\documentclass[a4paper]{article}

\usepackage[T1]{fontenc}
\usepackage{amsmath} 
\usepackage{amsthm}
\usepackage{amsfonts}
\usepackage{amssymb}
\usepackage[title]{appendix}
\usepackage{booktabs}

\usepackage{graphicx}

\usepackage[round]{natbib}
\author{Shin Irgens Banshoya\footnote{Corresponding author, shin.banshoya@uib.no} \footnote{Center for modeling of Coupled Subsurface Dynamics, Department of Mathematics, University of Bergen, Allegaten 41, Bergen, 5020, Norway} \quad Inga Berre\textsuperscript{\textdagger} \quad Eirik Keilegavlen\textsuperscript{\textdagger}
}

\title{A simulation study of the impact of fracture networks on the co-production of geothermal energy and lithium}
\date{}

\begin{document}

\maketitle

\begin{abstract}
Co-production of geothermal energy and lithium is an emerging opportunity with the potential to enhance the economic potential of geothermal operations. 
The economic reward of extracting lithium from geothermal brine is determined by how the lithium concentration evolves during brine production.
In the initial stage, production will target lithium contained in the brine resident close to the production well.
While lithium recharge, in the form of rock dissolution and inflow from other parts of the reservoir, is possible, the efficiency of such recharge depends on the geology of the reservoir.
In this work, we study how structural heterogeneities in the form of fractures impact the flow of lithium-carrying brine.
Using a numerical simulation tool that gives high resolution of flow and transport in fractures and the host rock, we study how the presence of fractures influences energy and lithium production.
Our simulations show that, due to heat conduction and the lack of mineral recharge from the rock, differences in fracture network geometries have a much larger impact on lithium production than energy production.
The simulations thus confirm that in addition to the geochemical characterisation of lithium in geothermal brines, understanding fracture characterisation and its impact on production is highly important for lithium production.
\end{abstract}

\section{Introduction}
The need for lithium is growing exponentially, driven by the need for rechargeable lithium batteries. As global demand for lithium is expected to increase by a factor of 40 from 2020 to 2040 in the IEA Sustainable Development Scenario \citep{IEA2021}, securing a reliable and diversified supply of lithium has become a top priority for many nations. 

The substantial presence of lithium in geothermal brines presents an opportunity for the co-production of geothermal energy and lithium. For example, in both the US and Europe, several geothermal regions have lithium-rich brines, creating the potential for a domestic lithium supply chain \citep{SANJUAN2022102385,stringfellow2020retrospective}. 

Operating geothermal systems typically involve a combination of injection and production wells. Re-injection is essential for disposing of the cooled produced fluids and maintaining the reservoir pressure. However, in many cases, this process leads to the breakthrough of re-injected fluids \citep{gringarten1978reservoir, stefansson1997geothermal,li2016thermal}. As a result, progressively higher concentrations of re-injected fluids are produced. Over time, this leads to lower temperatures of the produced fluid. In geothermal-lithium co-production, lithium production rates are expected to decline over time due to the chemical breakthrough of lithium-depleted re-injection fluid. While the rate at which lithium leaches from the rock may be too low to significantly increase the lithium concentration in the re-injected fluid \citep{jungmannEtAl2024}, conductive heat transfer between rock and fluid and between warmer and colder regions will contribute to the heating of the injected fluid. This leads to an earlier chemical breakthrough compared to thermal breakthrough \citep{goldberg2023challenges}. 

The timing of chemical and thermal breakthroughs, along with subsequent declines in lithium and heat production, is influenced by the flow field in the reservoir \citep{dobson2023characterizing, goldberg2023challenges}. The flow field is primarily determined by the pressure gradients induced through injection and production combined with the permeability of the formation. Structural heterogeneity in permeability, resulting, for example, from features such as networks of faults and fractures, plays a crucial role in determining the flow in the system \cite{moreno1994flow, egert2020implications}. For example, if fractures form high-permeable pathways between the injection and production well, the injection fluid will reach the production well early, and the subsequent ratio of injected fluid that is produced will be large (flow-short circuiting) \citep{liu2020numerical,fadel2022causes}. In contrast, if fractures provide high-permeable pathways connecting the wells to other parts of the reservoir, this will hamper the breakthrough of injected fluid in the production well, and even after injected fluid has reached the production well, the ratio of injected fluid that is produced will be low. 

The large effect fractures and faults have on fluid flow also affects lithium and energy production, but the relative effect on lithium and energy production is not the same \citep{goldberg2023challenges}. At the pore scale, heat is transferred from the solid
rock to the colder injected fluids, while at the continuum scale, conduction transfers heat from high to low-temperature regions.
The large aspect ratio of fractures implies that 
conductive heat transfer from the surrounding porous formation will contribute significantly to heating the injected fluid. These heat transfer mechanisms dampen the effect of the convective heat transfer dictated by the flow field and, hence, the effect the fracture network geometry has on the thermal breakthrough. 
In contrast, lithium production is more dictated by the flow field as the concentration transport is mainly driven by advection \citep{goldberg2023challenges}. It will, therefore, to a higher degree, be influenced by fracture network geometry. In summary, models that include the effect of faults and fractures are needed to optimise re-injection schemes for the combined production of geothermal energy and lithium. 

In this study, we explore the influence of fractures on chemical and thermal breakthroughs in geothermal lithium co-production using a mathematical model and corresponding numerical approach that simulates heat transfer and solute transport in fractured porous media. Our goal is to investigate the hypothesis that differences in fracture network geometry have a larger impact on lithium production than energy production.

The model integrates fluid flow, convective and conductive heat transfer, and solute transport of lithium-depleted re-injection fluid within a fractured reservoir. Although water-rock interactions can potentially leach lithium from minerals into the reservoir fluid \citep{druppel2020experimental, regenspurg2016fluid}, experimental data necessary for accurately quantifying this effect in simulation models are lacking. Therefore, we adopt a conservative approach and assume that the leaching of lithium from rock minerals into the lithium-depleted re-injection fluid is negligible \citep{dobson2023characterizing}. The model is based on an explicit representation of fractures, which are treated as objects of one dimension lower than the surrounding medium \citep{martin2005modeling}. This mixed-dimensional model accurately represents the processes in the fractures and their interaction with matrix processes \citep{BERRE2021103759}.

The article is structured as follows. We present the mathematical model and numerical approach in Sec. \ref{sec:model}. We present the simulation results in Sec. \ref{sec:num_results}. Finally, we give conclusions in Sec. \ref{sec:final_remarks}.

\section{Mathematical model and numerical approach} \label{sec:model}
We consider a fractured porous media where, following \citep{martin2005modeling}, the fractures are treated as two-dimensional surfaces in three-dimensional domains and one-dimensional lines in two-dimensional domains. This section presents the governing model equations for flow and transport in the matrix and fractures, respectively, and the equations defining the coupling between the matrix and fractures. A short review of the numerical approach is given at the end. 

\subsection{Matrix equations} \label{sec:matrix_pde}
We consider incompressible single-phase flow, modelled by Darcy's law and the mass conservation equation

\begin{align}
    \mathbf{v}=- &\frac{\mathcal{K}}{\eta} \nabla p, \label{eq:matrix_darcy} \\
    \nabla \cdot \mathbf{v} &= f_{p}, \label{eq:matrix_mc}
\end{align}
where $\mathbf{v}$, $\mathcal{K}$, $\eta$ and $p$ are the Darcy flux, the permeability, the dynamic viscosity and the pressure, respectively.
Lastly, $f_{p}$ is the volumetric flow rate per unit bulk volume that accounts for the well injection and production. With $\mathfrak{q}$ denoting a constant flow rate, it is defined as
\begin{equation*}
    f_{p}= 
    \begin{cases}
         \mathfrak{q}, \quad &\text{injection points,} \\
        -\mathfrak{q}, \quad &\text{production points,} \\
         0. \quad &\text{elsewhere.}
    \end{cases}
\end{equation*}

Neglecting dispersion, the non-reactive transport equation for the lithium concentration reads
\begin{equation}
    \partial(\phi c )+\nabla \cdot (c \mathbf{v})=f_{c}, \label{eq:solute_transport_matrix}
\end{equation}
where $\phi$ is the porosity, $c$ is the molar lithium concentration and
\begin{equation*}
    f_{c}= 
    \begin{cases}
         c_{\text{in}} \mathfrak{q}, \quad &\text{injection point,} \\
        -c \mathfrak{q}, \quad &\text{production point,} \\
         0, \quad &\text{elsewhere,}
    \end{cases}
\end{equation*}
the accounts for the injection and production of the lithium concentration, with $c_{\text{in}}$ being the injected concentration. 

The temperature, $T$, evolves due to convective heat flux and Fourier's law in an energy conservation equation. We further assume local thermal equilibrium and, hence, write the energy conservation equation as 
\begin{equation}
 \partial_t((\rho b)_{\mathfrak{m}} T) + \nabla \cdot ((\rho b)_{\mathfrak{f}} T \mathbf{v} -\kappa_{\mathfrak{m}}\nabla T)= f_{T}. \label{eq:matrix_temp}
\end{equation}
Here, $(\rho b)_{\mathfrak{m}}=\phi(\rho b)_{\mathfrak{f}} + (1-\phi)(\rho b)_{\mathfrak{s}}$ and $\kappa_{\mathfrak{m}}=\phi\kappa_{\mathfrak{f}} + (1-\phi)\kappa_{\mathfrak{s}}$ are the effective heat capacity and thermal conductivity, respectively, with the subscripts $\mathfrak{f}$ and $\mathfrak{s}$ referring to the fluid and solid rock. The density, the specific heat capacity and the thermal conductivity are denoted $\rho$, $b$ and $\kappa$, respectively, where the subscript indicates if they are a fluid or a solid rock parameter. These parameters are taken to be constant.  
Finally, $f_{T}$ represents the injection and production rate of the temperature,

\begin{equation*}
    f_{T}= 
    \begin{cases}
         (\rho b)_{\mathfrak{f}}T_{\text{in}}\mathfrak{q}, \quad &\text{injection point,} \\
        -(\rho b)_{\mathfrak{f}}T\mathfrak{q}, \quad &\text{production point,} \\
         0, \quad &\text{elsewhere,}
    \end{cases}
\end{equation*}
where $T_{\text{in}}$ is the injected temperature.

\subsection{Fracture equations} \label{sec:frac_pde}
This section describes the partial differential equations in fractures and fracture intersections. We use the modelling strategy of treating fractures as lower-dimensional objects embedded in the porous matrix, joint with interface conditions for the matrix-fracture interaction. 
Since this modelling approach has been presented in detail several times before (e.g. \citealp{BERRE2021103759, GLASER2022110715, keilegavlen2021porepy, frih2012modeling, budivsa2021block, HYMAN2022111396}, and references therein), we simplify the presentation given here by considering only a single fracture in a porous medium and refer to the mentioned references for the more general case with several fractures, as well as how this modelling strategy includes fracture intersections.

\begin{figure}[t]
    \centering
    \includegraphics[scale=0.45]{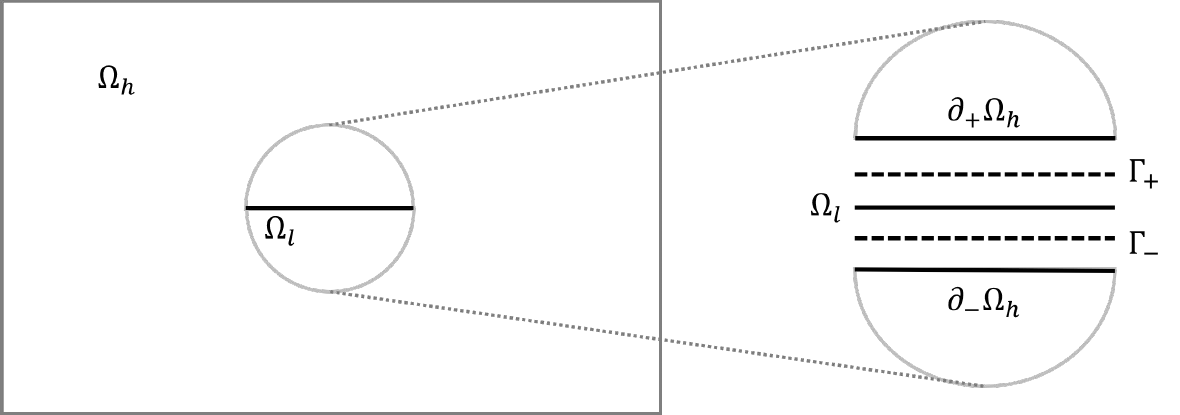} 
    \caption{Illustration of mixed-dimensional representation of a fractured porous medium with matrix $\Omega_{h}$ and one fracture $\Omega_{l}$. The internal boundaries of $\Omega_h$, $\partial_{+}\Omega_{h}$ and $\partial_{-}\Omega_{h}$, and the interfaces, $\Gamma_{+}$ and $\Gamma_{-}$, between $\Omega_{h}$ and $\Omega_{l}$ are shown to the right. 
    }
    \label{fig:md_domain}
\end{figure}

Following the domain decomposition of \cite{martin2005modeling} and \cite{boon_2018}, the $d$-dimensional fractured porous medium $\Omega$, $d=2,3$, is decomposed into two subdomains. The first is $\Omega_{h}$, the $d$-dimensional porous matrix subdomain, and the second is $\Omega_{l}$, the ($d-1$)-dimensional fracture subdomain. The two subdomains are connected by two interfaces, $\Gamma_{+}$ and $\Gamma_{-}$, and two internal matrix boundaries, $\partial_{+}\Omega_{h}$ and $\partial_{-}\Omega_{h}$. Fig. \ref{fig:md_domain} shows the connection between the matrix and the fracture. In the following, a variable with the subscript $h$ or $l$ belongs to $\Omega_{h}$ or $\Omega_{l}$, respectively. Similarly, a variable with the subscript $+$ or $-$ belongs to $\Gamma_{+}$ or $\Gamma_{-}$.

The flow equations in the fracture read
\begin{align}
&\mathbf{v}_{l} = -\epsilon_{l} \frac{\mathcal{K}_{l}}{\eta_{l}} \nabla p_{l}, \label{eq:flow_darcy_frac} \\
\nabla \cdot \mathbf{v}_{l}& = \epsilon_{l}f_{p} + (\mathrm{v}_{+} + \mathrm{v}_{-}) |_{\Omega_{l}}. \label{eq:flow_mc_frac}
\end{align}
In these equations, the scaling by the aperture, $\epsilon_{l}$, accounts for the dimension reduction. The aperture also governs the fracture permeability via the cubic law, 
\begin{equation*}
    \mathcal{K}_{l}=\epsilon_{l}^{2}/12. 
\end{equation*}
Furthermore, the nabla operator and the fracture permeability operate in the tangential direction of the fracture. 
Lastly, $\mathrm{v}_{+}$ and $\mathrm{v}_{-}$ on the right-hand side in Eq. ($\ref{eq:flow_mc_frac}$) represent interface fluxes that account for the fracture flow coupling to the matrix flow. They are modelled by the Darcy-type equation \citep{martin2005modeling}
\begin{equation}
    \mathrm{v}_{j}=-\frac{2\mathcal{K}_{l}|_{\Gamma_{j}}}{(\eta \epsilon)_{l} |_{\Gamma_{j}} }\big(p_{l}|_{\Gamma_{j}} - p_{h}|_{\Gamma_{j}} \big),
\end{equation}
where the subscript $j$ refers to either the $+$ or $-$ side of the fracture. 

The matrix flow Eqs. ($\ref{eq:matrix_darcy}$)-($\ref{eq:matrix_mc}$) are coupled to the fracture flow equations through an internal Neumann condition
\begin{equation}
    (\boldsymbol{\nu}_{h} \cdot \mathbf{v}_{h}) |_{\partial_{j}\Omega_{h}} = \mathrm{v}_{j}|_{\partial_{j}\Omega_{h}}.\label{eq:flow_neu_cond}
\end{equation}

The transport equation for the concentration in the fracture is
\begin{equation}
    \partial(\epsilon_{l} \phi_{l} c_{l})+\nabla \cdot (c_{l} \mathbf{v}_{l}) = \epsilon_{l}f_{c} + (\zeta_{+}+\zeta_{-})|_{\Omega_{l}}. \label{eq:solute_transport_frac}
\end{equation}
The advective interface flux, $\zeta_{j}$, is calculated as
\begin{equation}
    \zeta_{j}=\begin{cases}
    \mathrm{v}_{j} c_{h}|_{\Gamma_{j}}, & \text{if $\mathrm{v}_j \geq 0,$} \\
    \mathrm{v}_{j} c_{l}|_{\Gamma_{j}}, &\text{if $\mathrm{v}_j < 0,$}
    \end{cases}
\end{equation}
and appears in the following Neumann condition for $\Omega_h$
\begin{equation}
     (\boldsymbol{\nu}_{h} \cdot c_{h} \mathbf{v}_{h})\rvert_{\partial_j\Omega_h} = \zeta_{j}|_{\partial_{j}\Omega_{h}}.  \label{eq:transport_interface_neumann_cond}
\end{equation}

Finally, the energy conservation equation in the fracture is
\begin{equation}
\begin{split}
 \partial_t( \epsilon_{l} (\rho b)_{\mathfrak{m},l} T) + &\nabla \cdot ((\rho b)_{f,l} T_{l}\mathbf{v}_{l} - \epsilon_{l} \kappa_{\mathfrak{m},l}\nabla T_{l})=  \\
 &\epsilon_{l}f_{T} + \{(w_{+}+q_{+})+(w_{-}+q_{-})\}|_{\Omega_{l}}, \label{eq:frac_temp}
\end{split}
\end{equation}
with the internal boundary conditions 
\begin{alignat}{3}
(\boldsymbol{\nu}_h \cdot (\rho b)_{\mathfrak{f},h}T_{h}\mathbf{v}_{h}) \rvert_{\partial_j\Omega_h} &= w_j|_{\partial_{j}\Omega_{h}}, \\
(\boldsymbol{\nu}_h \cdot -\kappa_{\mathfrak{m},h}\nabla T_{h})\rvert_{\partial_j\Omega_h} &= q_{j}|_{\partial_{j}\Omega{h} }.
\end{alignat}
The interface convective and conductive fluxes, $w_j$ and $q_j$, are calculated as
\begin{align} 
    w_j&=\begin{cases} \mathrm{v}_{j} ((\rho b)_{\mathfrak{f},h} T_{h})|_{\Gamma_{j} }, &\text{ if } \mathrm{v}_{j} \geq 0, \\
    \mathrm{v}_{j}((\rho b)_{\mathfrak{f},l} T_{l})|_{\Gamma_{j}} , &\text{ if } \mathrm{v}_{j} <0,
    \end{cases} \\
    q_j & =-\frac{2 \kappa_{\mathfrak{f},l}|_{\Gamma_{j} } } { \epsilon_{l} |_{\Omega_{l}} } (T_{l}|_{\Gamma_{j}} - T_{h}|_{\Gamma{j}}).\label{eq:interface_flux_heat}
\end{align}

\subsection{Numerical approach} \label{sec:numerical_strategy}

Since the flow field will be constant based on the choice of boundary conditions and source term, the flow Eqs. ($\ref{eq:matrix_darcy}$)-($\ref{eq:matrix_mc}$), ($\ref{eq:flow_darcy_frac}$)-($\ref{eq:flow_neu_cond}$) are solved once at the beginning of the simulation to obtain the flow field used to solve for temperature and lithium concentration in time.
Then, in the time loop, the remaining equations (Eqs. (\ref{eq:solute_transport_matrix})-(\ref{eq:matrix_temp}), (\ref{eq:solute_transport_frac})-(\ref{eq:interface_flux_heat})) are solved.

The equations are discretised and solved using functionality from the open-source software PorePy \citep{keilegavlen2021porepy}.
The simulation meshes in PorePy are constructed to conform to fractures in accordance with the modelling principles employed in Section \ref{sec:frac_pde}, with the actual mesh generation handled by a Gmsh backend \citep{geuzaine2009gmsh}. 
For the spatial discretisation, we use the cell-centred finite volume methods implemented in PorePy. 
Specifically, a two-point flux approximation is used for the elliptic terms, and the standard first-order upwind scheme is used for the advective terms. The interface fluxes are calculated using the subdomain variables evaluated at the internal boundaries. 
For temporal discretisation, we employ the implicit Euler scheme. 
The full discretisation of the unsteady PDEs leads to a system of linear equations, which is solved by the solver PyPardiso \citep{haas2023pypardiso}.

\section{Simulations of evolution of lithium concentration and temperature} \label{sec:num_results}
This section presents simulations that investigate the effect of fractures on lithium concentration and temperature in the reservoir during production, as well as their effect on the produced lithium concentration and temperature. 
We start by considering stochastically generated fracture networks for different fracture densities in a two-dimensional domain.
We then consider a three-dimensional test case that illustrates the effect of fracture network connectivity on lithium concentration and temperature for a more realistic setup in terms of fracture network geometry. 
The fluid and rock parameters are constant for all the simulations and are listed in Table \ref{tab:constant_simulation_parameters}.   

\begin{table}[t]
    \centering
    \begin{minipage}{224pt}
    \caption{Constant simulation parameters for the simulations in Sections \ref{sec:stoch_generated_networks} and \ref{sec:three_dim_simulation}.}
    \begin{tabular}{lc}
    \toprule
    Parameter & Value \\
    \midrule
    Dynamic viscosity ($\eta$) & $10^{-3}$ Pa s \\
    Matrix porosity & $0.2$ $[-]$ \\
    Fracture porosity & $1.0$ $[-]$ \\
    Matrix permeability ($\mathcal{K}$) & $1.0 \cdot 10^{-13}$ m$^{2}$ \\
    Fracture aperture ($\epsilon$) & $5 \cdot 10^{-3}$ m \\
    Fluid density ($\rho_{\mathfrak{f}}$) & $1000$ kg/m$^{3}$ \\
    Solid density ($\rho_{\mathfrak{s}}$) & $2750$ kg/m$^{3}$ \\
    Fluid specific heat capacity ($b_{\mathfrak{f}}$) & $4200$ J/(kg K) \\
    Solid specific heat capacity ($b_{\mathfrak{s}}$) & $790$ J/(kg K) \\
    Fluid thermal conductivity ($\kappa_{\mathfrak{f}}$) & $0.6$ W/(m K) \\
    Solid thermal conductivity ($\kappa_{\mathfrak{s}}$) & $3.0$ W/(m K) \\
    \bottomrule
    \end{tabular}
    \label{tab:constant_simulation_parameters}
    \end{minipage}
\end{table}

\subsection{Stochastically generated fracture networks in two dimensions}\label{sec:stoch_generated_networks}
To study the impact of fractures on lithium and energy production, we consider stochastically generated fracture networks in a two-dimensional domain with different fracture densities. The two-dimensional domain can be considered as a horizontal slice of unit thickness taken out of a $3d$ formation with vertical fractures.

\subsubsection{Setup}
The computational domain is $\Omega = [0,3] \text{ km} \times [0,2]$ km, with an injection well at ($1,\, 1$) km and a production well at ($2,\, 1$) km.
We assume that the fracture centres, $(x_{c}, y_{c})$, follow a uniform distribution 
\begin{align*}
    x_{c} &\sim U([0,\, 3000]), \\
    y_{c} &\sim U([0,\, 2000]).
\end{align*}
Furthermore, the fracture length follows a lognormal distribution, with a mean of $500$ m and a standard deviation of $200$ m. The orientation is generated from a uniform distribution, $U[-\pi/2, \,\pi /2]$. 

\begin{figure}[t]
    \centering
    \includegraphics[scale=0.30]{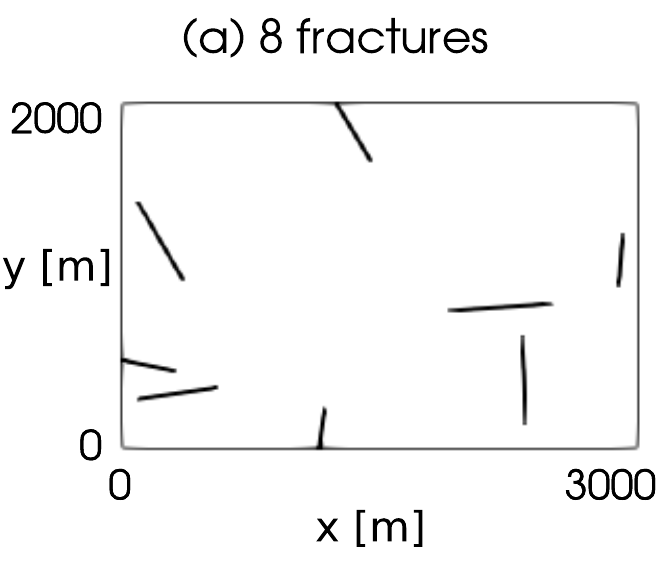}
    \includegraphics[scale=0.30]{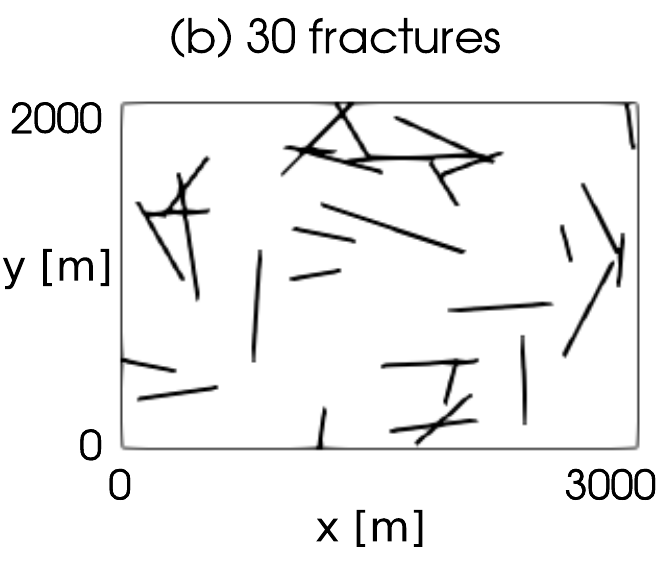}
    \includegraphics[scale=0.30]{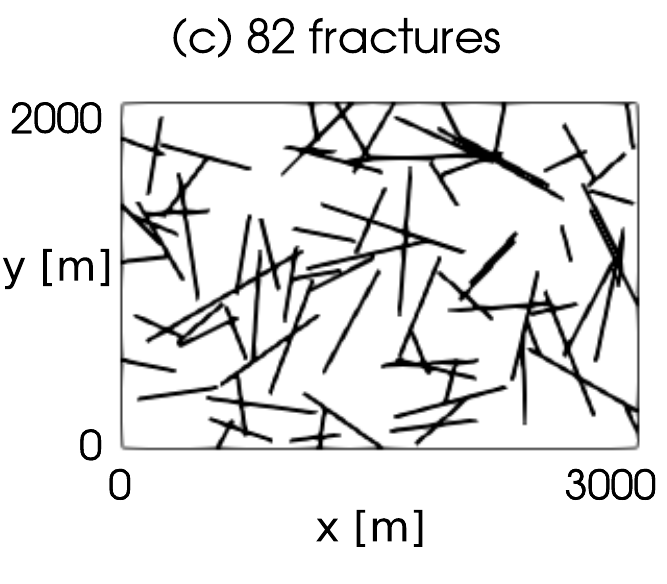}
    \caption{Example of stochastically generated fracture networks with (a) $8$, (b) $30$ and (c) $82$ fractures.}
    \label{fig:examples_of_generated_networks}
\end{figure}

We use the number of fractures to measure the fracture network density and consider three cases with $8$, $30$ and $82$ fractures in the domain, respectively. Fig. \ref{fig:examples_of_generated_networks} shows an example of a generated fracture network for each density.
The lowest fracture density is considered to represent a case where the fractures have a low probability of forming high-permeable connected pathways between the wells.
In the highest considered fracture density, the fracture network will, with a very high probability, provide long, high-permeable pathways for the fluid flow and thus completely dominate the transport of energy and lithium.

Before production, the reservoir is assumed to have an initial temperature of $160^{\circ}$C and an initial lithium concentration of $170$ mg/L.
For pressure, we set the constant value of $30$ MPa at the boundary of the domain, which will allow for fluid flow in and out of the simulation domain due to pressure differences induced by the injection and production. The injection temperature is $70^{\circ}$C, and the lithium concentration of the injected fluid is $0$ mg/L. The injection and production rate is $\mathfrak{q} = 0.4$ L/s, corresponding to a rate of $40$ L/s for a formation with a thickness of 100 m.
Lastly, the time step is set constant to $4.73\cdot 10^{7}$ s ($1.5$ years).  

%For the fracture network generation, we consider three fracture densities using the P$_{21}$ criterium \citep{dershowitz_generate_frac_network}, i.e. number of fractures per area. 

\subsubsection{Results}
To exploit the effect of fracture network geometry on the spatial evolution of lithium concentration and temperature in the reservoir, we first discuss the results for two fracture network realisations with 82 fractures, R1 and R2. 
The network R1 is meshed with 13239 matrix and 1434 fracture cells, and the network R2 is meshed with 14881 matrix and 1602 fracture cells.

\begin{figure}[t]
    \centering
    \includegraphics[scale=0.4]{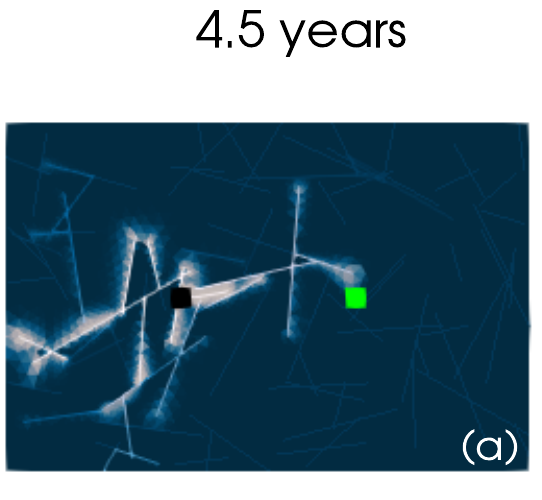}  \includegraphics[scale=0.4]{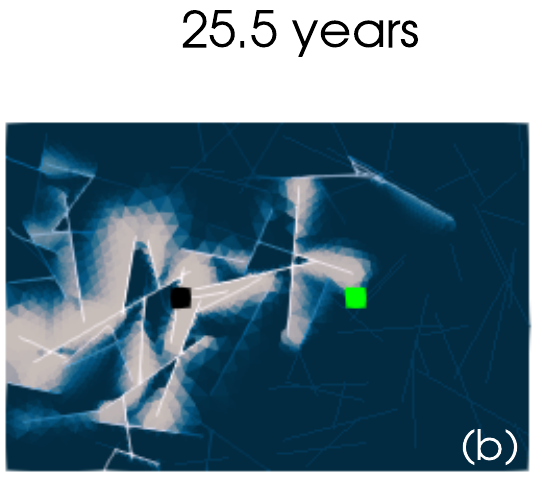}
    \includegraphics[scale=0.4]{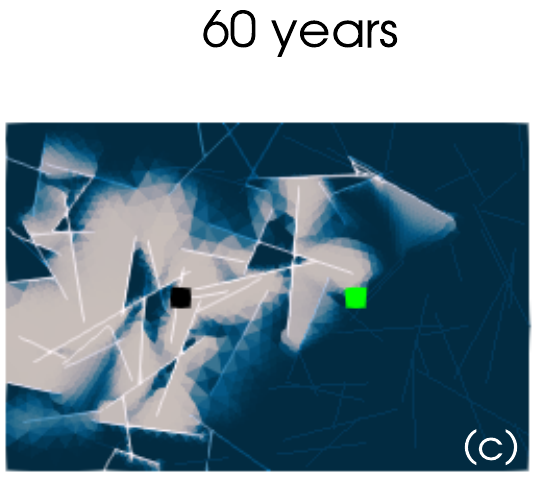} \\
    \includegraphics[scale=0.3]{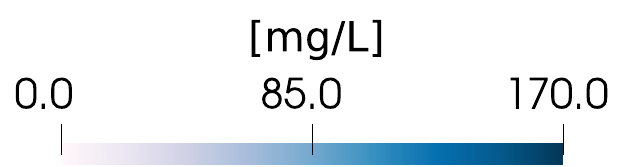} \\ \vspace{0.1cm}
    \includegraphics[scale=0.4]{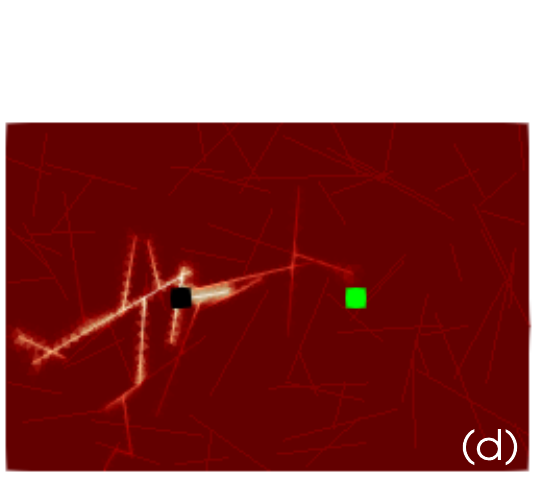}
    \includegraphics[scale=0.4]{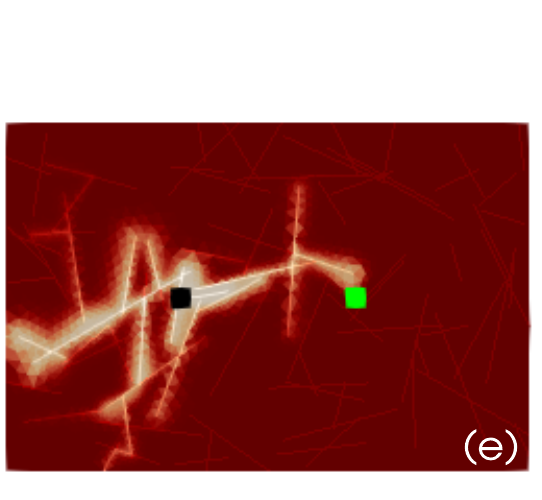}
    \includegraphics[scale=0.4]{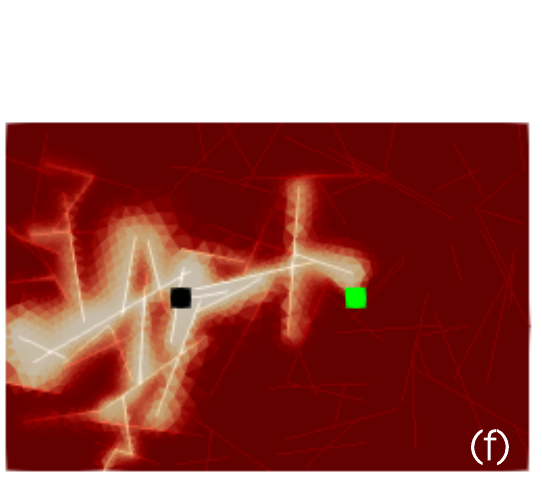} 
    \includegraphics[scale=0.3]{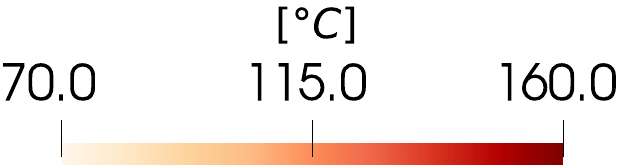}
    \caption{The spatial solution in the network R1 at different times. Lithium concentration after (a) $4.5$, (b) $25.5$ and (c) $60$  years, and temperature after (d) $4.5$, (e) $25.5$ years and (f) $60$ years. The black and green squares represent the injection and production wells, respectively.}
    \label{fig:spatial_evolution_105_fracs_R1}
\end{figure}

\begin{figure}[t]
    \centering
   \includegraphics[scale=0.4]{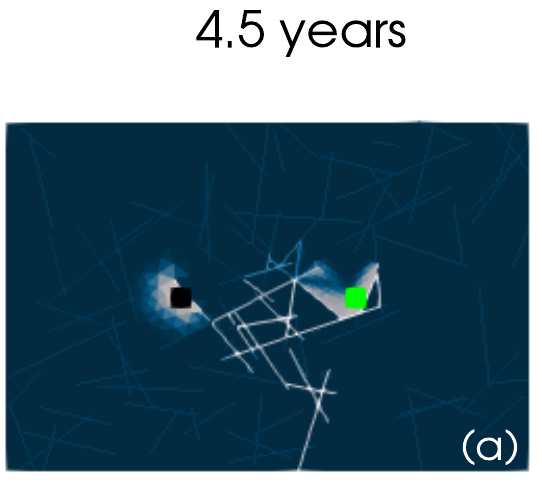}
    \includegraphics[scale=0.4]{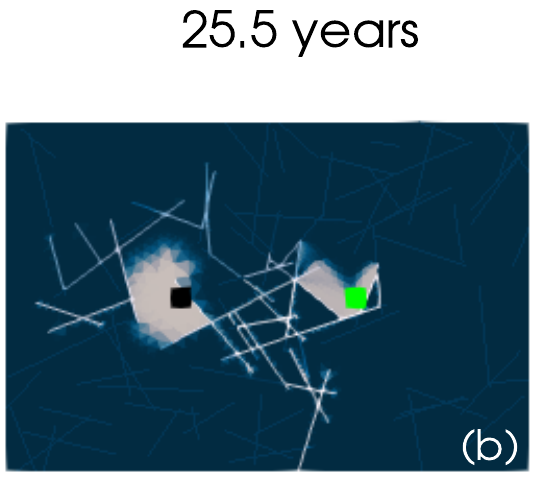}
    \includegraphics[scale=0.4]{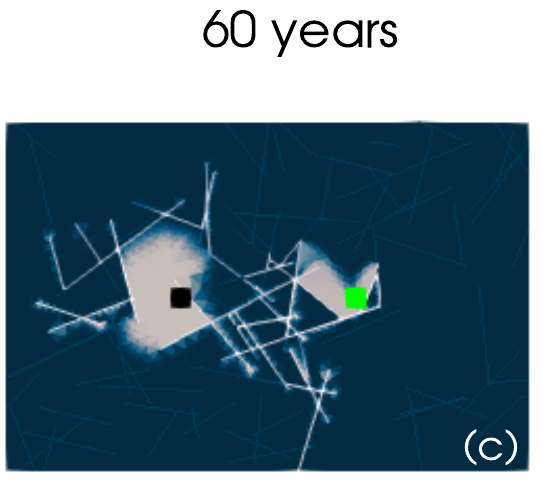} \\
    \includegraphics[scale=0.3]{Fig3_colbar_conc.pdf} \\ \vspace{0.1cm}
    \includegraphics[scale=0.4]{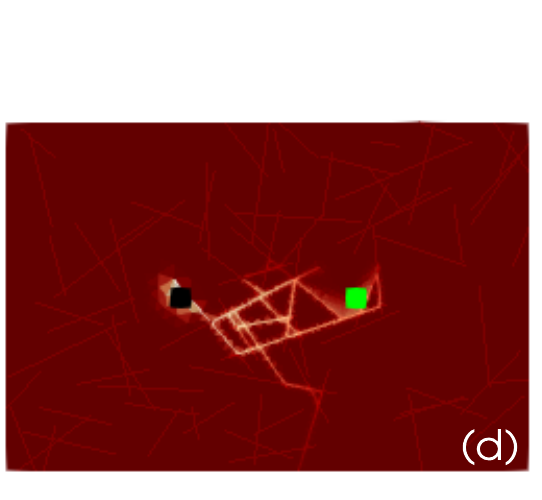}
    \includegraphics[scale=0.4]{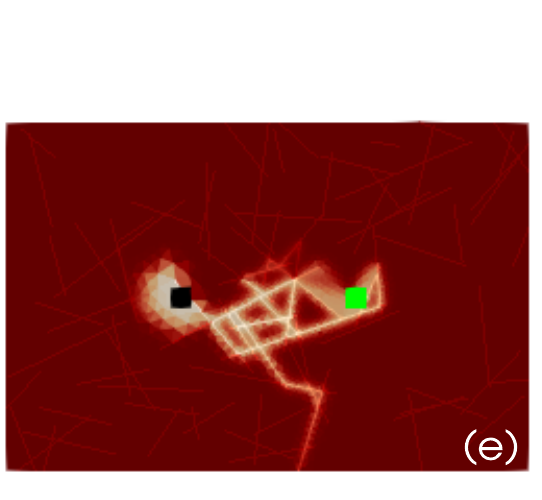}
    \includegraphics[scale=0.4]{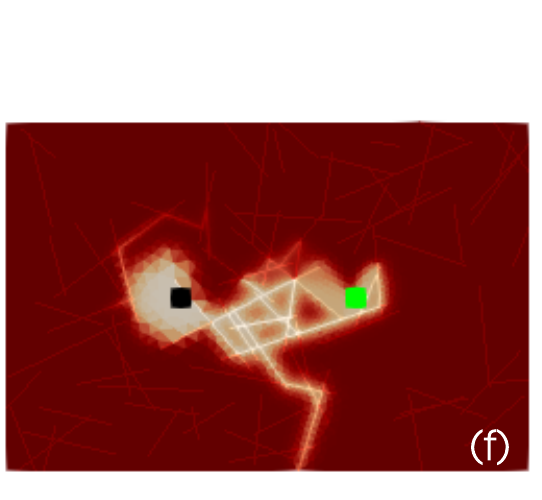} \\
    \includegraphics[scale=0.3]{Fig3_colbar_temp.pdf}
    
    \caption{The spatial solution in the network R2 at different times. Lithium concentration after (a) $4.5$,  (b) $25.5$ and (c) $60$  years, and temperature after (d) $4.5$, (e) $25.5$ years and (f) $60$ years. The black and green squares represent the injection and production wells, respectively.}
    \label{fig:spatial_evolution_105_fracs_R2}
\end{figure}

\begin{figure}[t]
    \centering
    \includegraphics[scale=0.25]{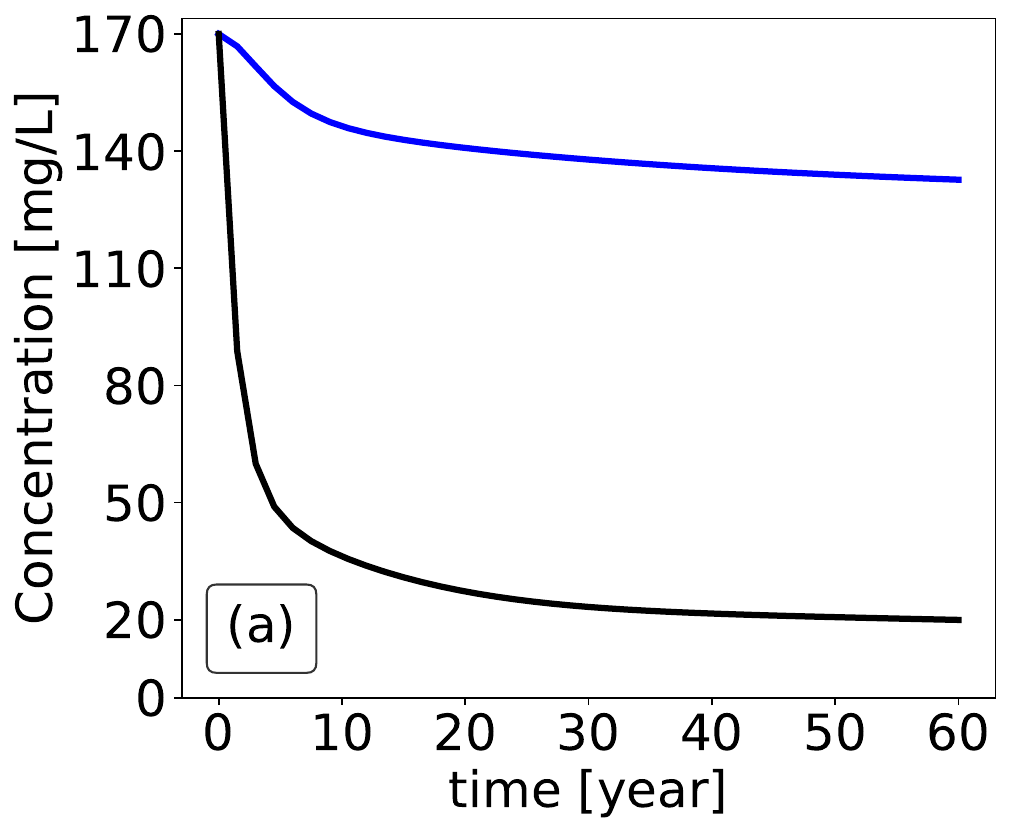}
    \hspace{0.3cm}
    \includegraphics[scale=0.25]{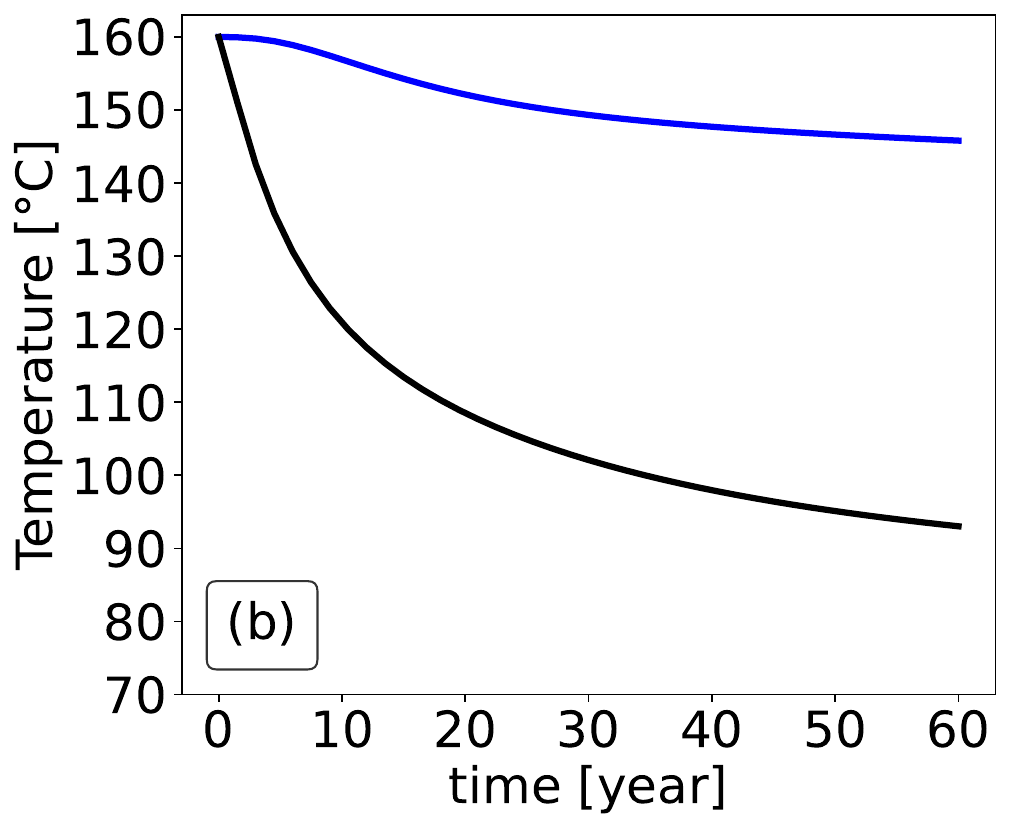} \\
    \includegraphics[scale=0.35]{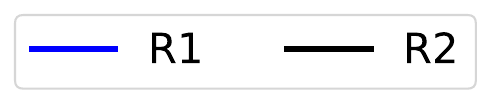}
    \caption{Lithium concentration (a) and temperature (b) in the production well as a function of time for R1 and R2.}
    \label{fig:subset_temp_mc_105_realisation}
\end{figure}

\begin{figure}[t]
    \centering
     \includegraphics[scale=0.25]{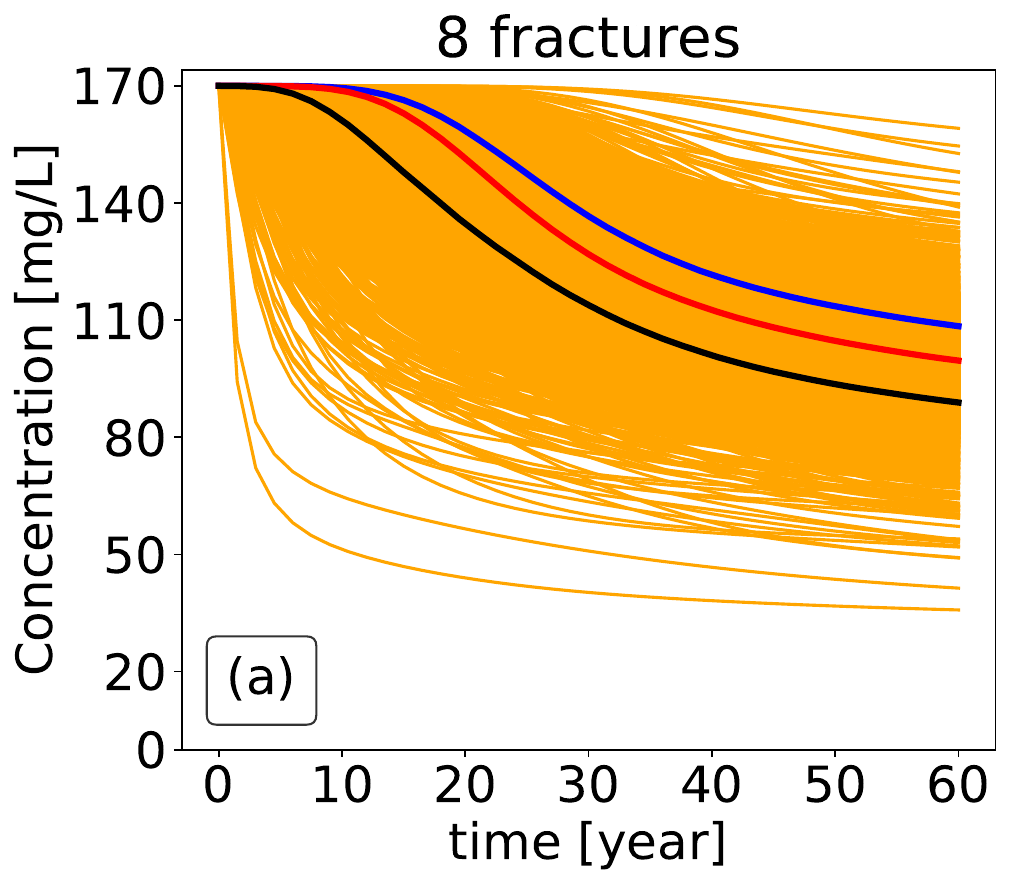}
     \hspace{0.3cm}
    \includegraphics[scale=0.25]{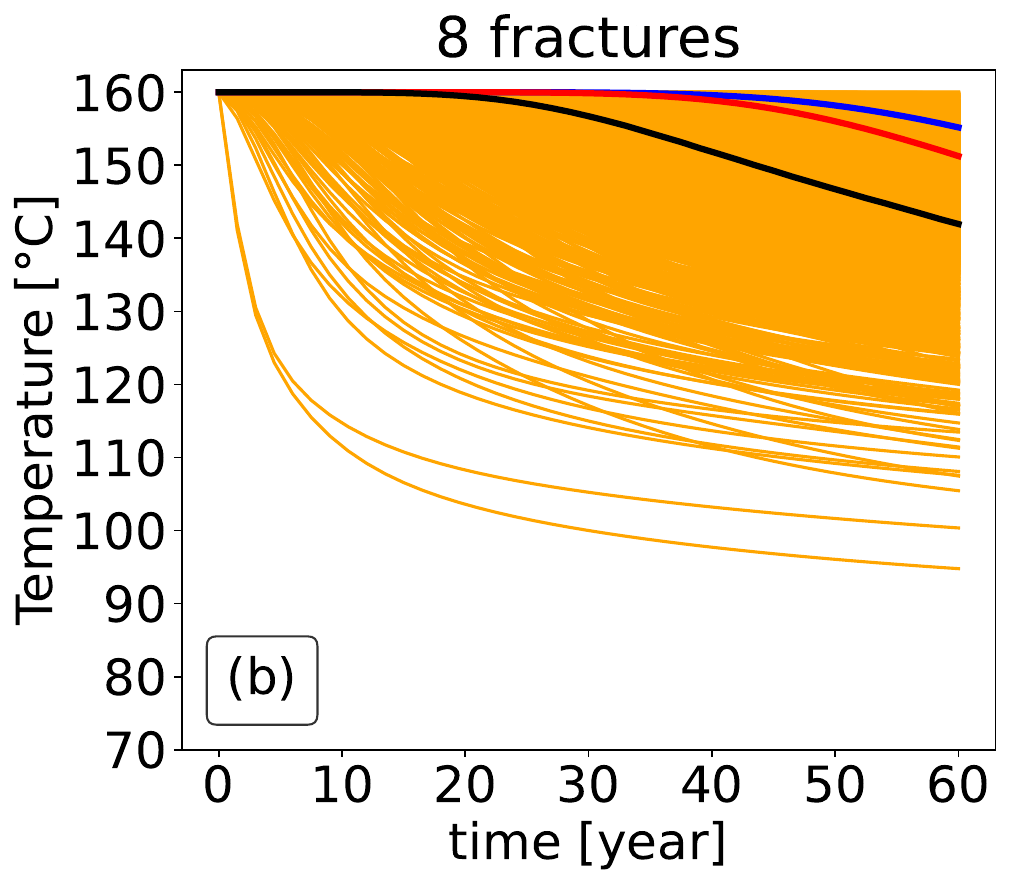} \vspace{0.2cm} \\
    \includegraphics[scale=0.25]{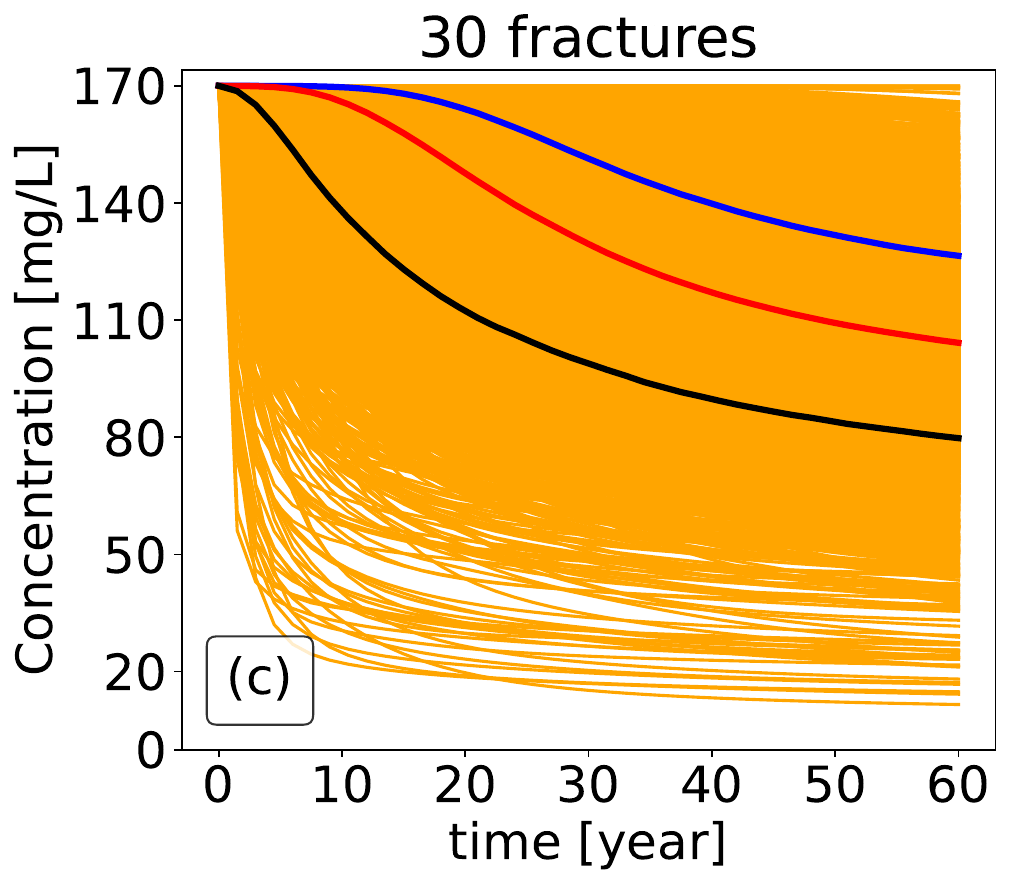}
     \hspace{0.3cm}
    \includegraphics[scale=0.25]{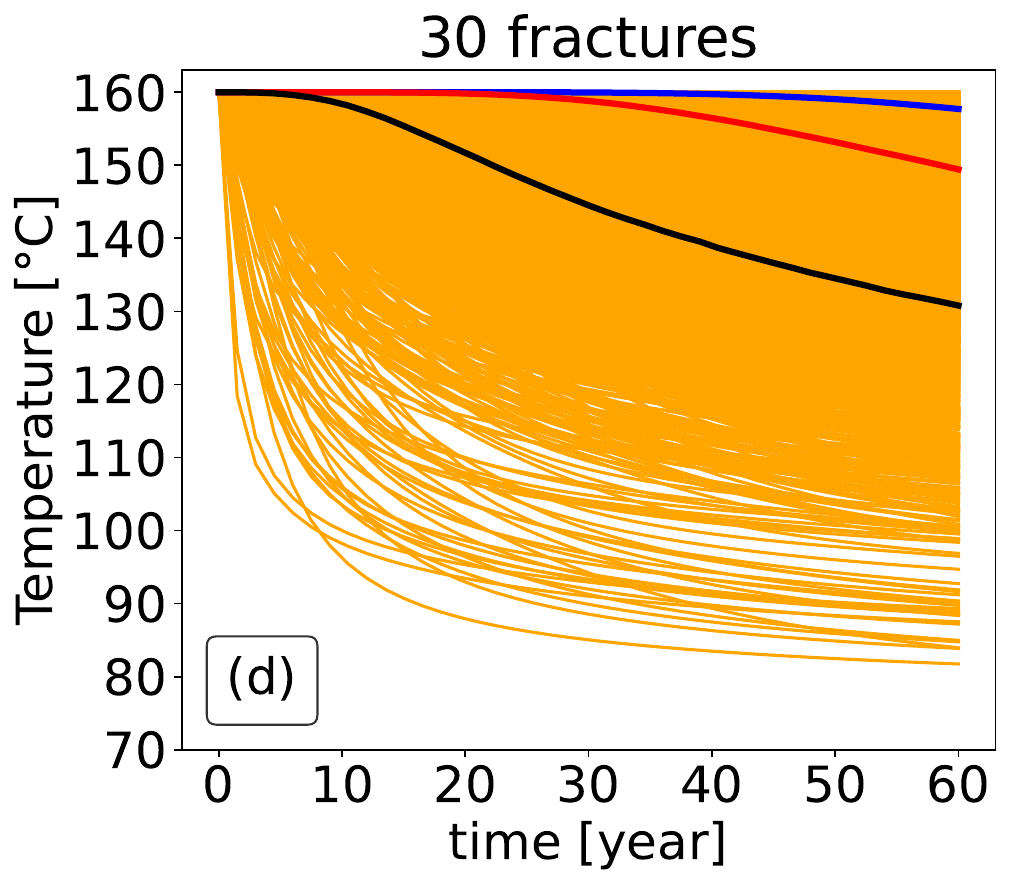}  \vspace{0.2cm}\\  
    \includegraphics[scale=0.25]{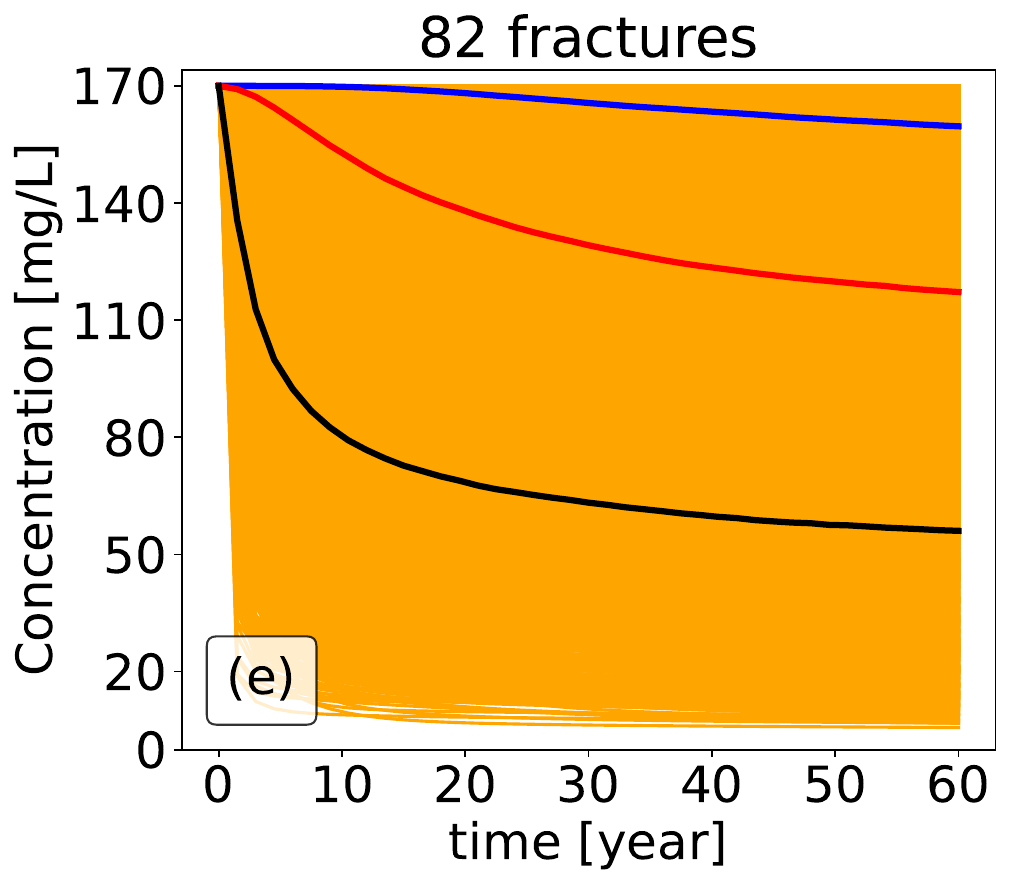}
     \hspace{0.3cm}
    \includegraphics[scale=0.25]{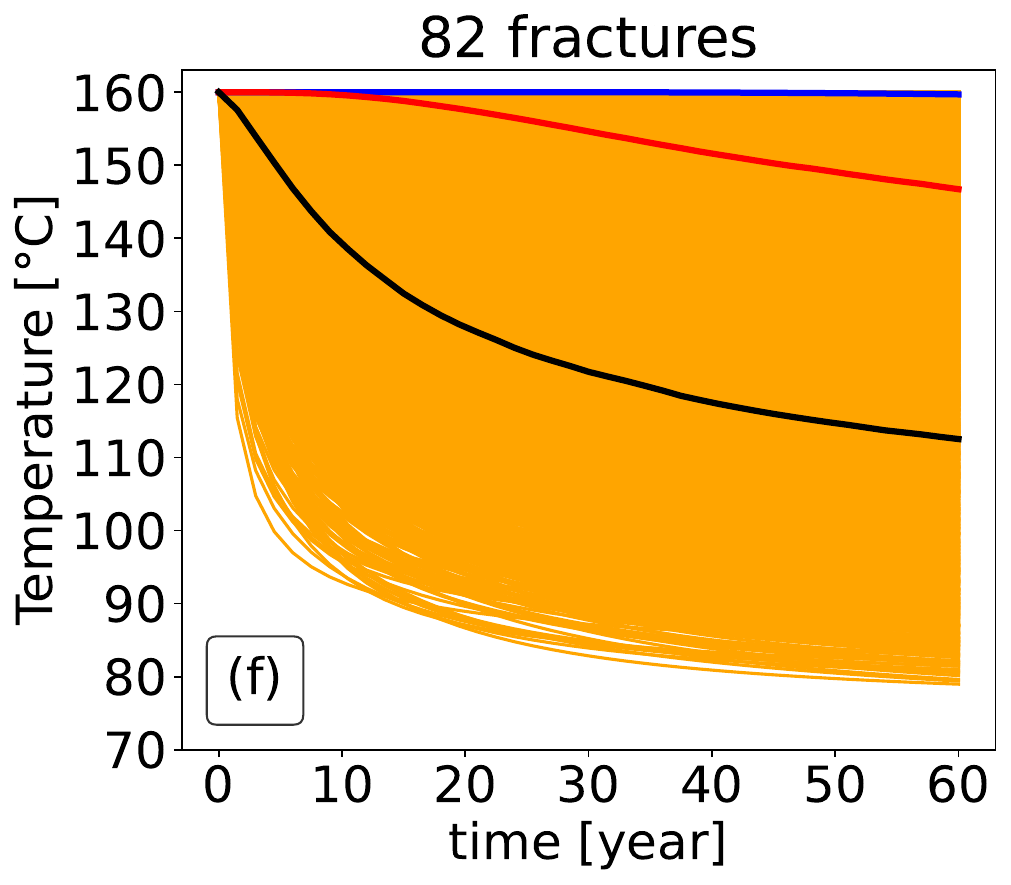} \\
    \includegraphics[scale=0.35]{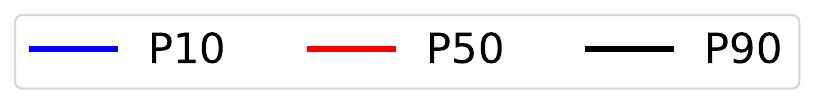}
    \caption{The 10,000 realisations (in orange) of the evolution of the lithium concentration (left column) and temperature (right column) in the production well, for fracture networks with $8$ ((a), (b)), $30$ ((c), (d)) and 82 ((e), (f)) fractures, joint with P$10$, P$50$ and P$90$ curves.}
    \label{fig:mc_realisation}
\end{figure}

Figs. \ref{fig:spatial_evolution_105_fracs_R1} and \ref{fig:spatial_evolution_105_fracs_R2} show the lithium concentration and temperature in the reservoir after $4.5$ years, $25.5$ years and $60$ years for R1 and R2, respectively.   
As can be seen from the figures, the lithium concentration and temperature evolve in the reservoir fundamentally different for the two different fracture networks: 
For R1, the fracture network provides preferential pathways for flow and transport towards the left boundary of the domain.
In contrast, for R2, the fracture network provides a preferential pathway between the injection and production wells, leading to flow short-circuiting between the wells. 
The result is very different spatial profiles for lithium concentration and temperature for the two fracture networks.
Though the figures also show some numerical diffusion, this does not significantly impact the production curves.

The difference in production between R1 and R2 is striking, as can be seen in Fig. \ref{fig:subset_temp_mc_105_realisation}, where the lithium concentration and temperature in the production well are shown as a function of time. 
For R2, the breakthrough of lithium-depleted injected fluid in the production well happens early. After approximately $10$ years, the decline in produced concentration is already almost $80\%$ of the maximum possible decline. In comparison, for the network R1, the decline in produced concentration is nearly $15\%$ after about $10$ years. 

Compared to the decline in concentration of produced lithium, the decline in production temperature is slower. 
This is expected, as heat is transferred from the solid rock to the fluids locally in the porous matrix, and since, in addition to convective heat transfer, conduction transfers heat from high to low-temperature regions.
After approximately $10$ years, the decline in produced temperature is around $2\%$ of the maximum possible decline for R1, while it is nearly $25\%$ for R2. 

From Fig. \ref{fig:subset_temp_mc_105_realisation}, when comparing the results obtained with R1 and R2, it is clear that the relative difference over time is larger in lithium production than in energy production. 
As this effect is consistent over time, the flow short-circuiting between the wells caused by the geometry of R2 will have a larger negative impact on the cumulative lithium production than energy production (relative difference of area under the graphs in Fig. \ref{fig:subset_temp_mc_105_realisation}).

\begin{figure}[t]
    \centering
     \includegraphics[scale=0.23]{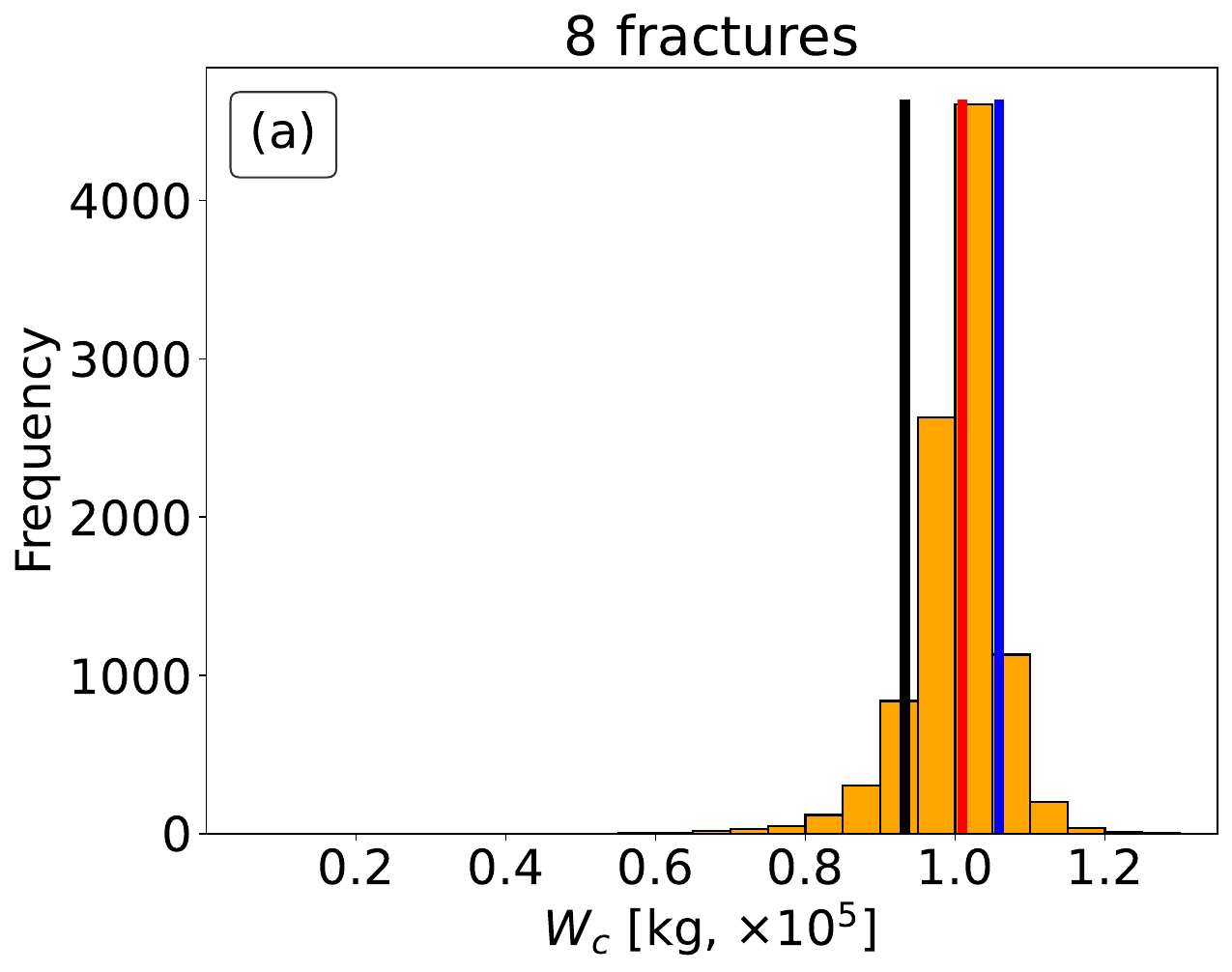}
     \hspace{0.3cm}
    \includegraphics[scale=0.23]{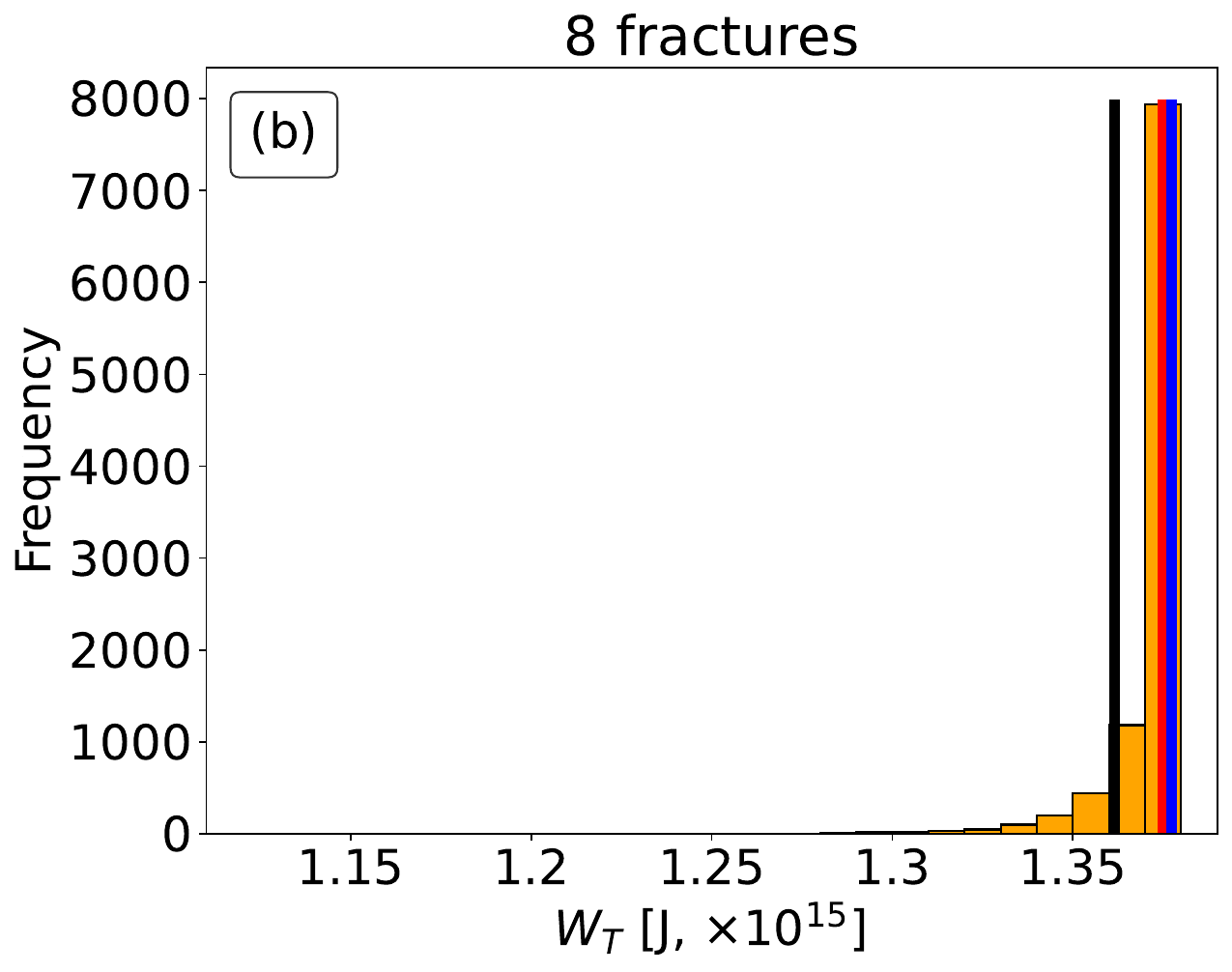} \vspace{0.2cm} \\
    \includegraphics[scale=0.23]{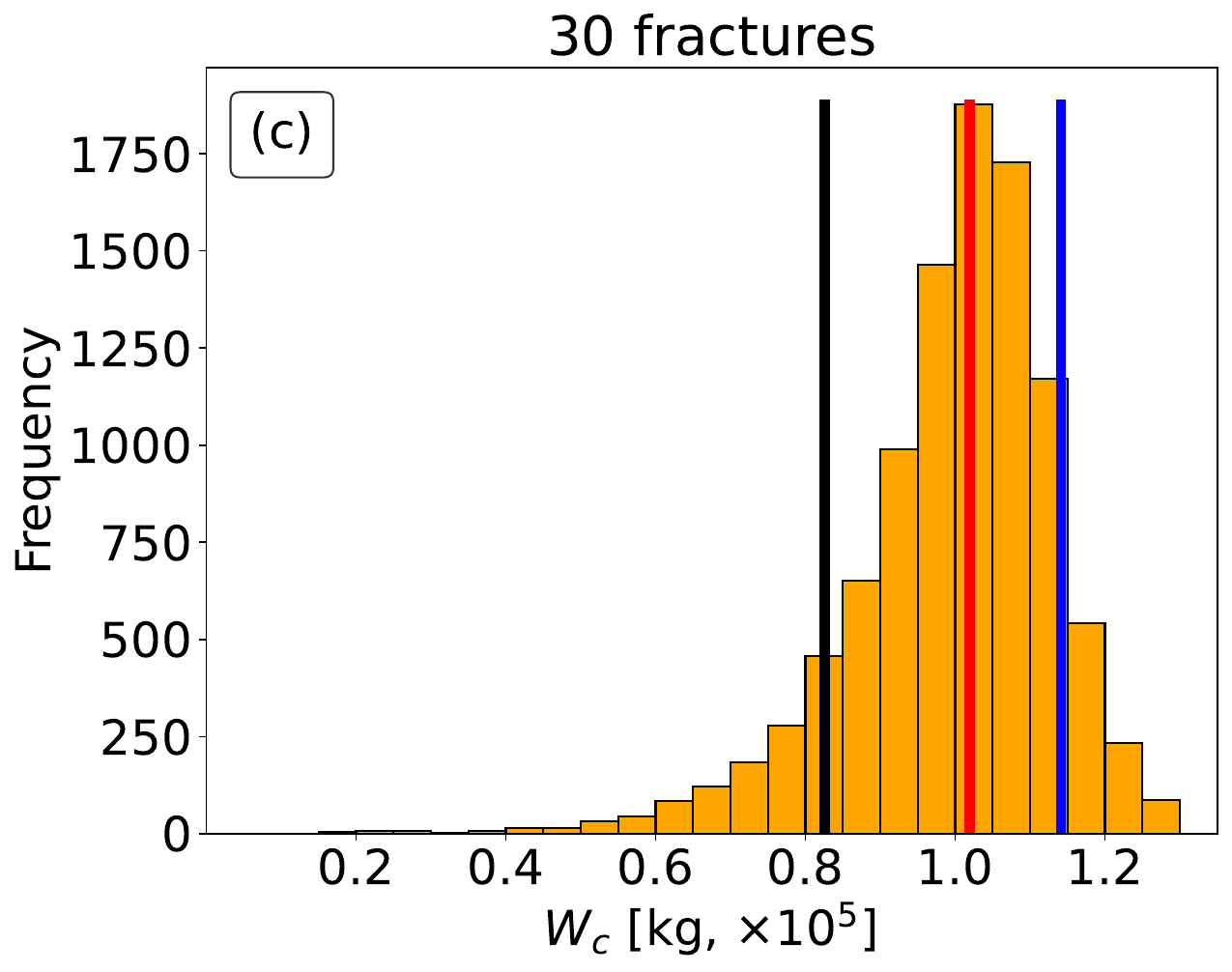}
     \hspace{0.3cm}
    \includegraphics[scale=0.23]{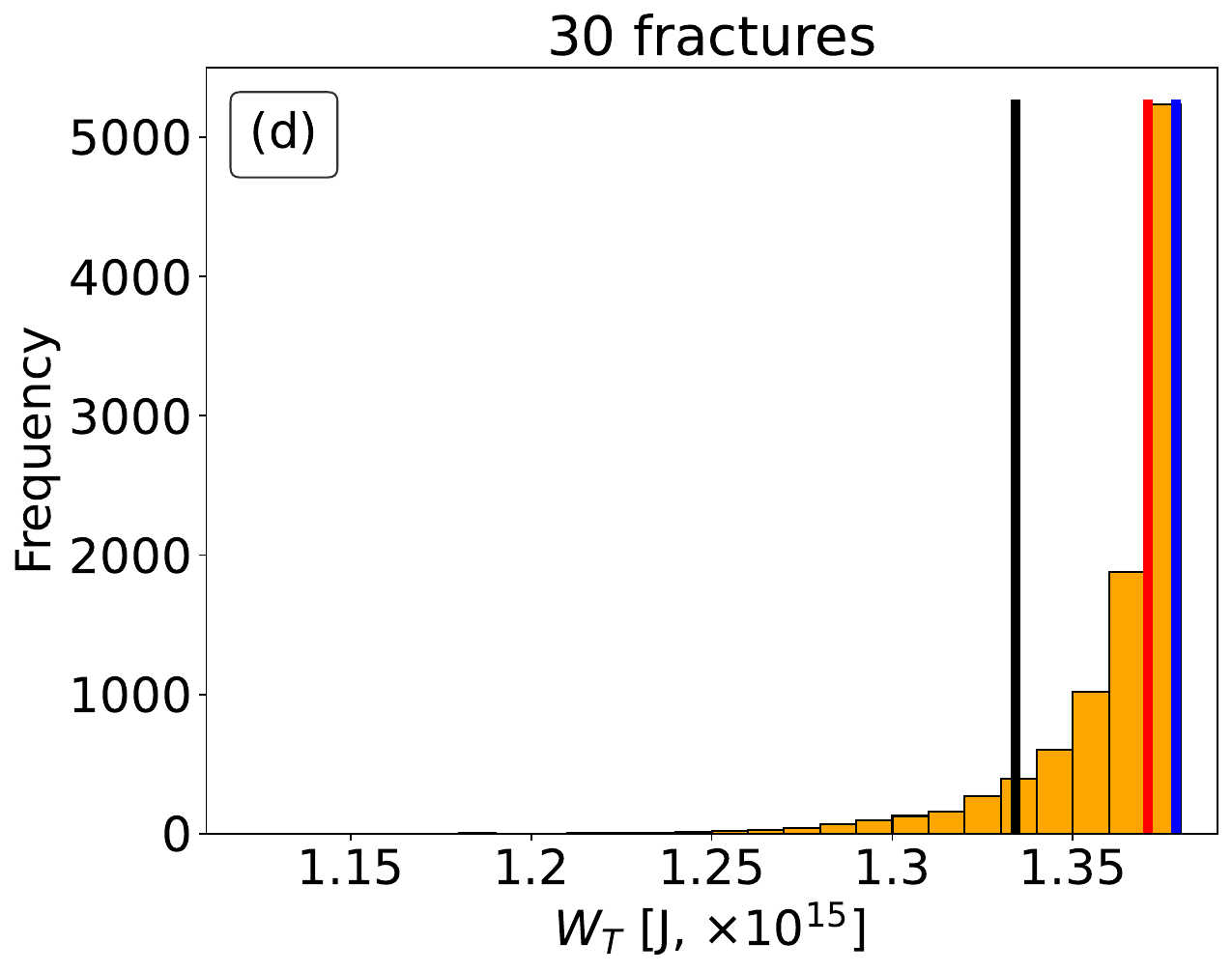}  \vspace{0.2cm}\\  
    \includegraphics[scale=0.23]{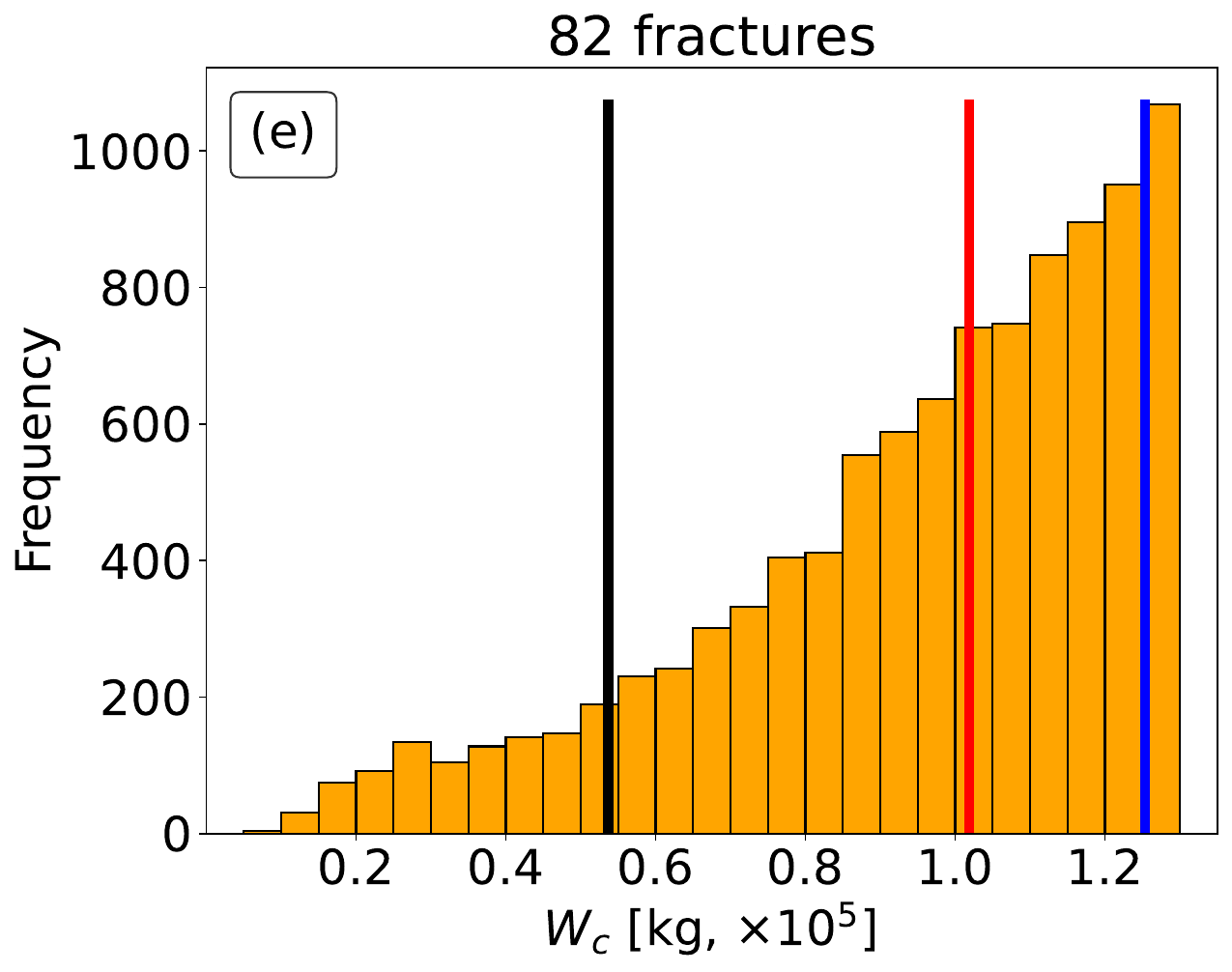}
     \hspace{0.3cm}
    \includegraphics[scale=0.25]{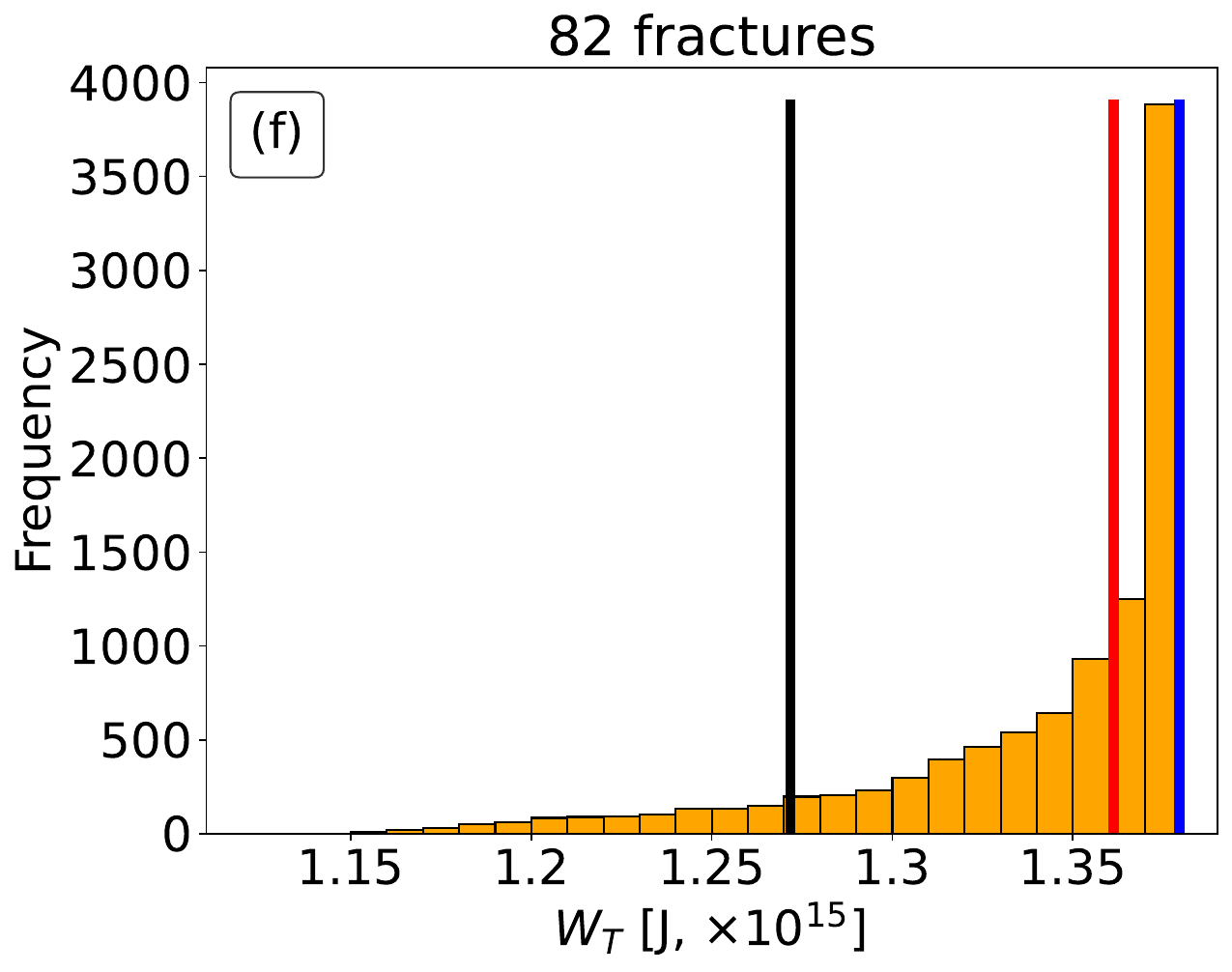} \\
    \includegraphics[scale=0.35]{Fig6_legend.pdf}
    \caption{Frequency histograms of the cumulative lithium ((a), (c), (e)) and energy ((b), (d), (f)) production for the different fracture densities.}
    \label{fig:mc_histograms}
\end{figure}
 
We next discuss the outcome of running $10,000$ realisations for each fracture density to further study the effects of fracture network geometries. 

Fig. \ref{fig:mc_realisation} shows the temporal evolution of concentration and temperature in the production cell, joint with the $10$th, $50$th and $90$th percentiles (P$10$, P$50$ and P$90$). 
For all fracture densities and P$10$, P$50$ and P$90$ curves, the front of low lithium concentration reaches the production well faster than the front of low temperature, as the latter experiences the combined effect of diffusion and the need to cool down the rock.
Notably, for the lithium concentration, the P$90$ curves show an almost immediate breakthrough for both the intermediate (30) and high (82) fracture densities; in the latter case, the P$50$ curve also indicates a rapid breakthrough.
By comparison,  a rapid breakthrough is only achieved for the P$90$ curve in the high-density regime for the cold temperature front.

We also see that increasing fracture density increases the likelihood (represented by the P$10$ curve) of a delayed breakthrough.
This can be attributed to the fracture network leading the injected fluid away from the production well. Although the limited domain size has some impact on the simulations, we believe this effect would also have been present for larger domains.
Nevertheless, the P$10$ curves also show that the lithium concentration in the production well decreases faster than the temperature. 

Fig. \ref{fig:mc_histograms} shows the frequency of the extracted lithium concentration, $W_{c}$, and temperature, $W_{T}$, accumulated over the production period, together with the associated P$10$, P$50$ and P$90$. 
The cumulative quantities are calculated as
\begin{equation*}
    W_{c}=\int_{0}^{60}c\mathfrak{q}\,dt, \qquad W_{T}=\int_{0}^{60}T\rho_{f} b_{f} \mathfrak{q}\,dt
\end{equation*}
The trapezoidal method is used to estimate the integrals.
By comparing lithium and energy production, it can immediately be seen that the relative span in the cumulative production is larger for lithium concentration than for temperature.
Comparing the P50 curves for the different fracture densities, we can see no large difference in total production in the period. However, the variation in the cumulative production is larger with an increasing number of fractures.    
Considering the P10 and P90 curves, over the 60-year production period, the difference between P10 and P90 in cumulative production is larger for lithium than for energy for all three fracture densities but increases with larger densities. 
As noted earlier, a high fracture density impacts production both positively and negatively, as the fracture networks either provide short pathways between the injection and production wells or spread the transported quantities more widely in the domain.  

In Appendix \ref{sec:appendix}, we demonstrate the convergence of the Monte Carlo simulations.

\subsection{Fracture networks in three-dimensions} \label{sec:three_dim_simulation} 

\begin{figure}
    \centering
    \includegraphics[scale=0.4]{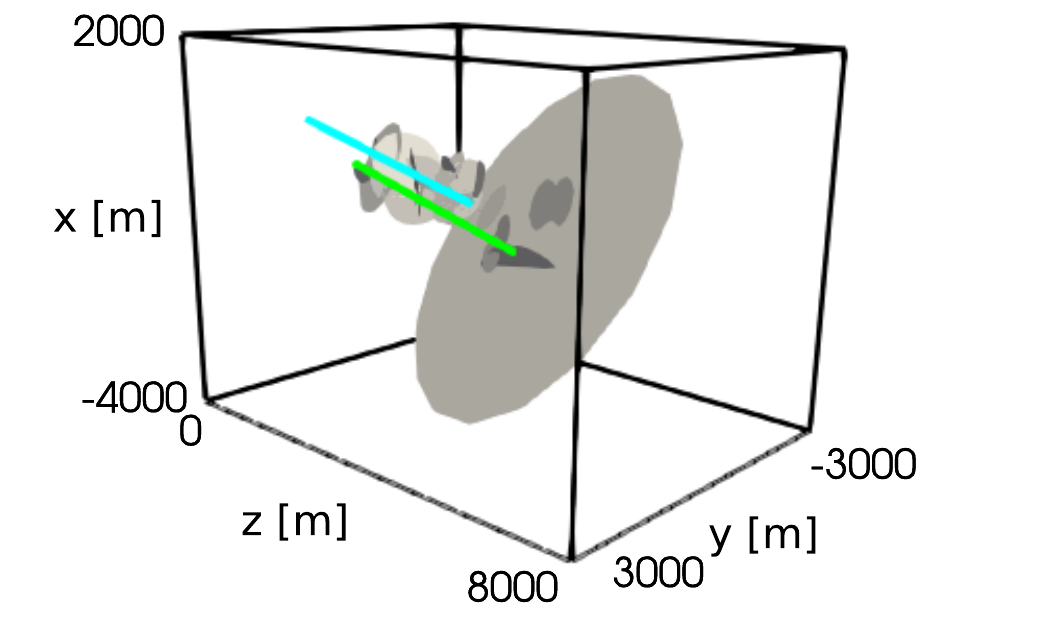}
    \caption{The fracture network N1 in the matrix. The black lines represent the global matrix boundaries, and the blue and green lines represent the injection and production well.}
    \label{fig:simulation_domain_3d}
\end{figure}

\begin{figure}[t]
    \centering
    \includegraphics[scale=0.4]{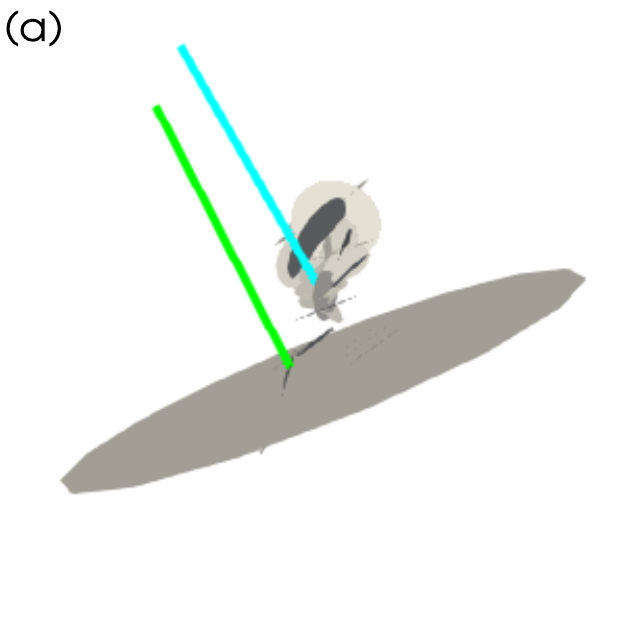} \hspace{1.0cm}
    \includegraphics[scale=0.4]{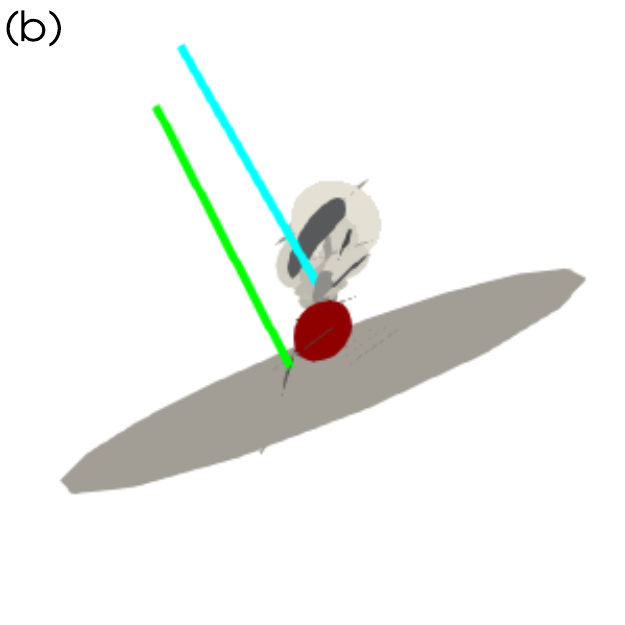}
    \caption{
     The fracture network N1 (a), where the fractures do not form a connected network between the injection and production well, and network N2 (b), where an additional fracture (illustrated in red) is added to the network N1.
    }
    \label{fig:without_and_with_extra_fracture}
\end{figure}

\begin{figure}[t]
    \centering
    \includegraphics[scale=0.4]{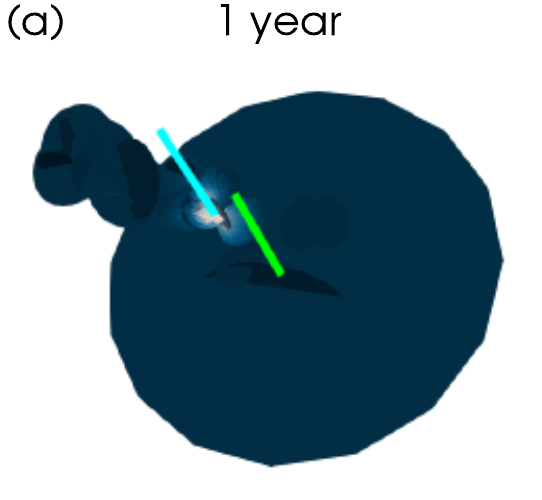} \hspace{1.0cm}
    \includegraphics[scale=0.4]{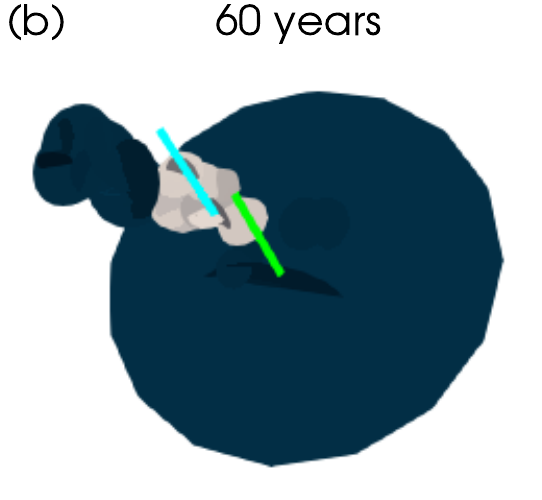} \\
    \includegraphics[scale=0.4]{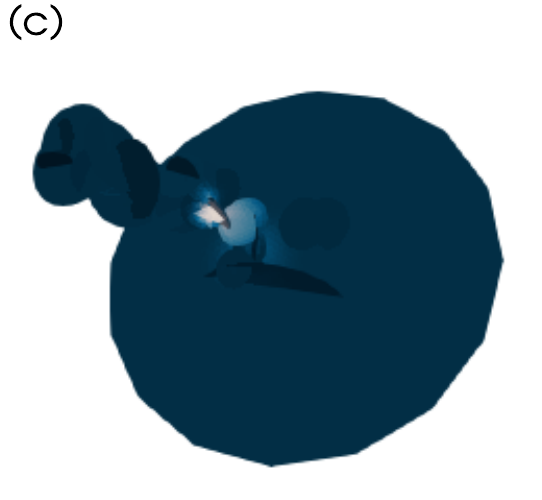} \hspace{1.0cm}
    \includegraphics[scale=0.4]{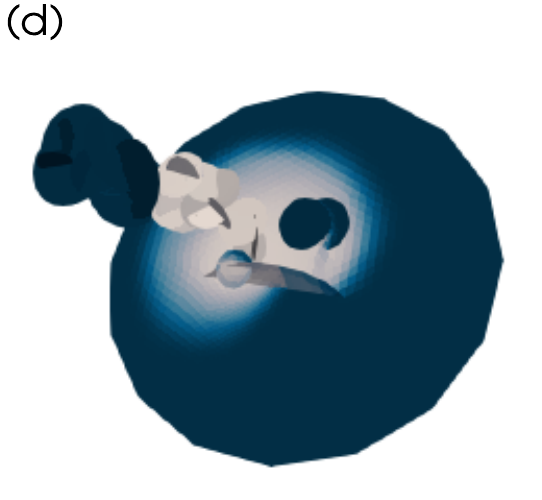} \\
    \includegraphics[scale=0.35]{Fig3_colbar_conc.pdf} \\ \vspace{0.3cm}
    \includegraphics[scale=0.4]{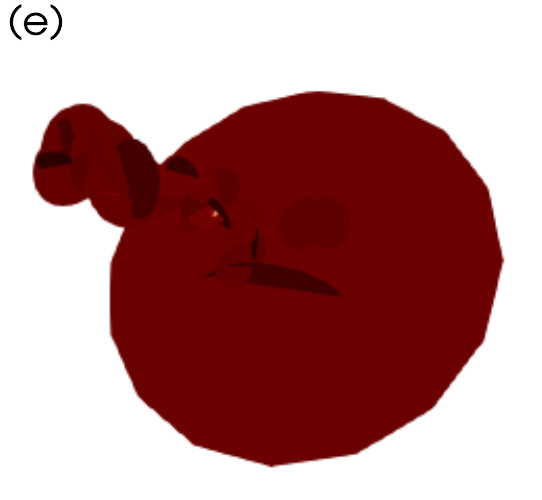} \hspace{1.0cm}
     \includegraphics[scale=0.4]{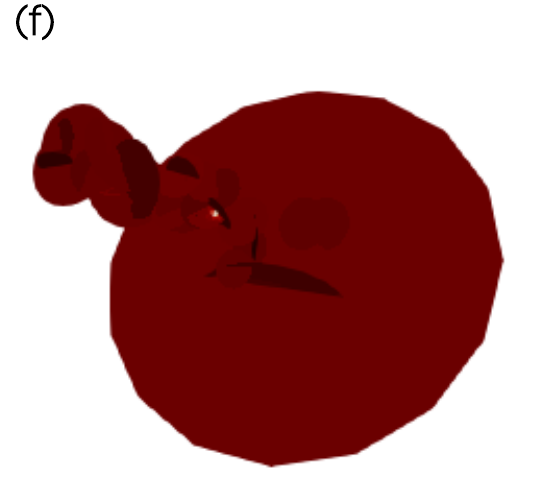} \\
    \includegraphics[scale=0.30]{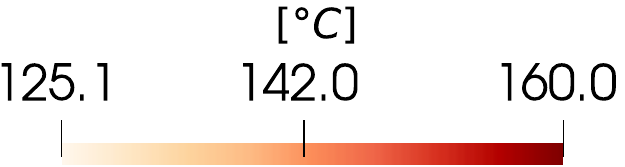} \hspace{1.5cm} 
    \includegraphics[scale=0.30]{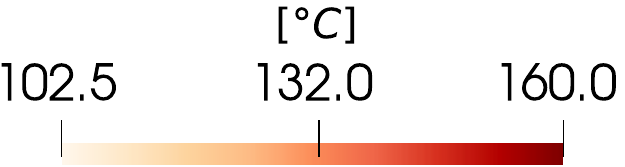} \\
    \caption{
   The simulation results after $1$ year (left column) and $60$ years (right column). (a) and (b) exhibit the lithium concentration in N1. (c) and (d) exhibit the lithium concentration in N2. (e) and (f) exhibit the temperature in N2.
    }
    \label{fig:res_without_extra}
\end{figure}

\begin{figure}[t]
    \centering
   \includegraphics[scale=0.28]{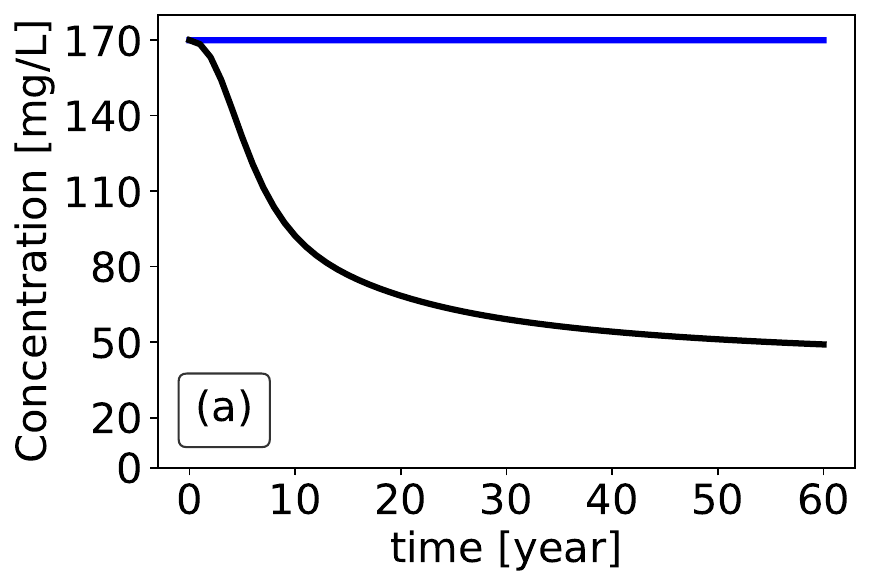} 
   \hspace{0.3cm}
   \includegraphics[scale=0.28]{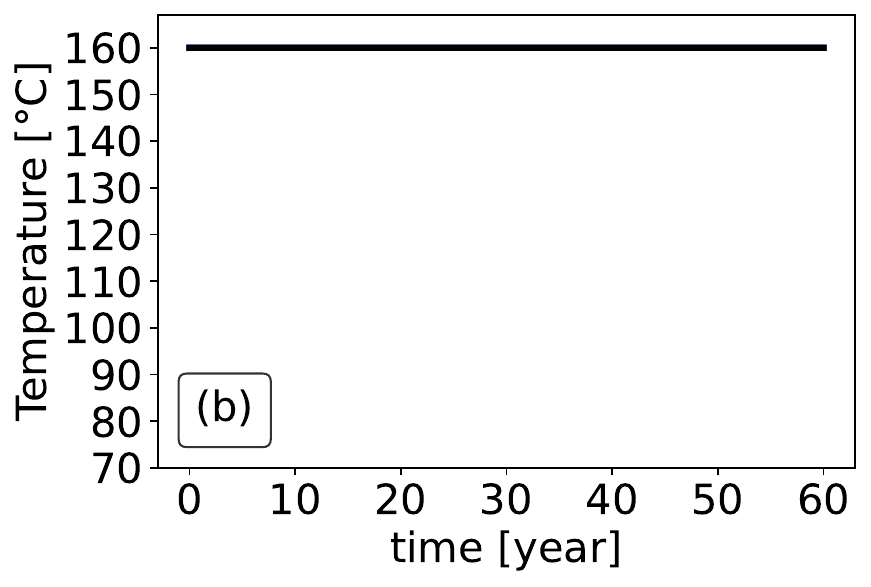} \\
   \includegraphics[scale=0.35]{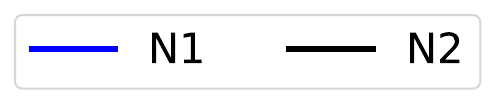}
   
    \caption{
     The lithium concentration (a) and temperature (b) 
     evolution at the production cell for the simulation with the fracture networks N1 and N2. 
    }
    \label{fig:produced_results_3d}
\end{figure}

In this section, we consider a three-dimensional fractured domain to exploit the characteristics identified for the two-dimensional simulations in a more realistic fracture network geometry. The fracture network geometry is inspired by information from the Soultz-sous For{\^e}ts geothermal site \citep{sausse20103d}. 
The network (N1) consists of $39$ fractures; a subset of these is illustrated in Fig. \ref{fig:simulation_domain_3d}, which also shows the injection and production well, respectively. 
The simulation domain is 
$\Omega = [-4, 2] \,\text{km} \times [-3, 3] \,\text{km} \times [0, 8] \, \text{km}$.

For the fracture network N1, the fractures do not form a connected network between the wells; see Fig. \ref{fig:without_and_with_extra_fracture}a. To investigate the effect of connectivity, we consider a second fracture network, N2, where we have introduced an additional fracture to the network N1 that links the disconnected parts of the N1 fracture network; see Fig. \ref{fig:without_and_with_extra_fracture}(b).

For the simulation with the fracture network N1, the matrix is meshed with 87072 cells, and the fractures are meshed with a total of 11032 cells.  
For the simulation with N2, the matrix is meshed with 86208 cells, and the fractures are meshed with a total of 11213 cells. The flow rate in the injection and production well is set to $\mathfrak{q}=30$ L/s, and the pressure is $30$ MPa at all global boundaries.
For the lithium concentration and temperature, the initial conditions, injection and boundary values are the same as in Section \ref{sec:stoch_generated_networks}. The time step is set constant to $3.15\cdot10^{7}$ s ($1$ year).

Fig. \ref{fig:res_without_extra} shows simulation results in the fracture networks N1 and N2 after $1$ year and $60$ years. 
As seen, the lithium concentration enters the computational domain from the injection well and propagates in the surrounding fracture network. 
For the simulation with the fracture network N1, where the fractures do not provide a direct connection between the injection and production cells, Figs. \ref{fig:res_without_extra}a-b show that the front of low concentration does not reach the production well. In contrast, for the simulation with the fracture network N2, Figs. \ref{fig:res_without_extra}c-d shows that the extra fracture creates a highly conductive pathway directly from the injection to the production well, which results in the breakthrough of the lithium-depleted fluid in the production well.

For the temperature, its front does not reach the production well for either of the fracture networks. For N2, Figs. \ref{fig:res_without_extra}e-f shows that the front of temperature does not migrate past the fracture where the injection well is located.
This is caused by the heat conduction into the injection fracture and the corresponding cooling of the rock matrix near the injection well.  

Fig \ref{fig:produced_results_3d}a displays the evolution of the concentration at the production well. In accordance with the spatial distribution of the lithium concentration, no reduction in the extracted lithium concentration is observed with N1. In contrast, for N2, the extracted lithium concentration is reduced from an early simulation time. 
For the energy production, no reduction in produced temperature is observed for either of the fracture networks, as observed in Fig. \ref{fig:produced_results_3d}b.
The example thus further underlines the importance of fracture connectivity and the different impact this has on energy and lithium production.

\section{Conclusion} \label{sec:final_remarks}

In this paper, we have numerically investigated the influence of fractures on the co-production of lithium and energy for the case of a geothermal doublet system where produced water is re-injected into the reservoir.

We have studied the difference in lithium concentration and temperature production from two-dimensional simulations with stochastically generated fracture geometries for different fracture densities. 
The simulations show that the impact of fractures is different for energy and lithium production, and the behaviour expected for energy production cannot readily be transferred to lithium production. 
Breakthrough of re-injected water causes a decline in both lithium concentration and temperature of the produced fluid. As expected, the decline in the lithium concentration always starts earlier than the decline in the temperature.
 
Over time, the relative reduction, compared to the case where the re-injected water does not reach the production well, will always be larger for cumulative lithium production than for energy production.
In terms of cumulative production, lithium production shows larger variation with different fracture geometries than energy production.
Hence, our results corroborate the hypothesis that differences in fracture network geometry have a larger impact on lithium production than energy production.

In the geothermal community, the importance of fractures and fracture network connectivity on energy production from a doublet configuration with injection and production well is well known and has been studied extensively. This paper shows that the impact of fractures and variety in fracture network geometry is even more important considering lithium production. Hence, modelling studies that appropriately account for the effect of fractures are crucial in estimating lithium production. 

\section*{Acknowledgements}

This work was financed by the Norwegian Research Council grant number 308733. The second author also acknowledges funding from the VISTA program, The Norwegian Academy of Science and Letters and Equinor.

\begin{appendices}

\section{} \label{sec:appendix}

In this appendix, we demonstrate the convergence of the Monte Carlo simulations in Sec. \ref{sec:stoch_generated_networks}
To assess the convergence, we have used an approach similar to that used by \cite{WANG2023102773} and \cite{CREMON2020107094}.

\begin{figure}[t]
    \centering
    \includegraphics[scale=0.26]{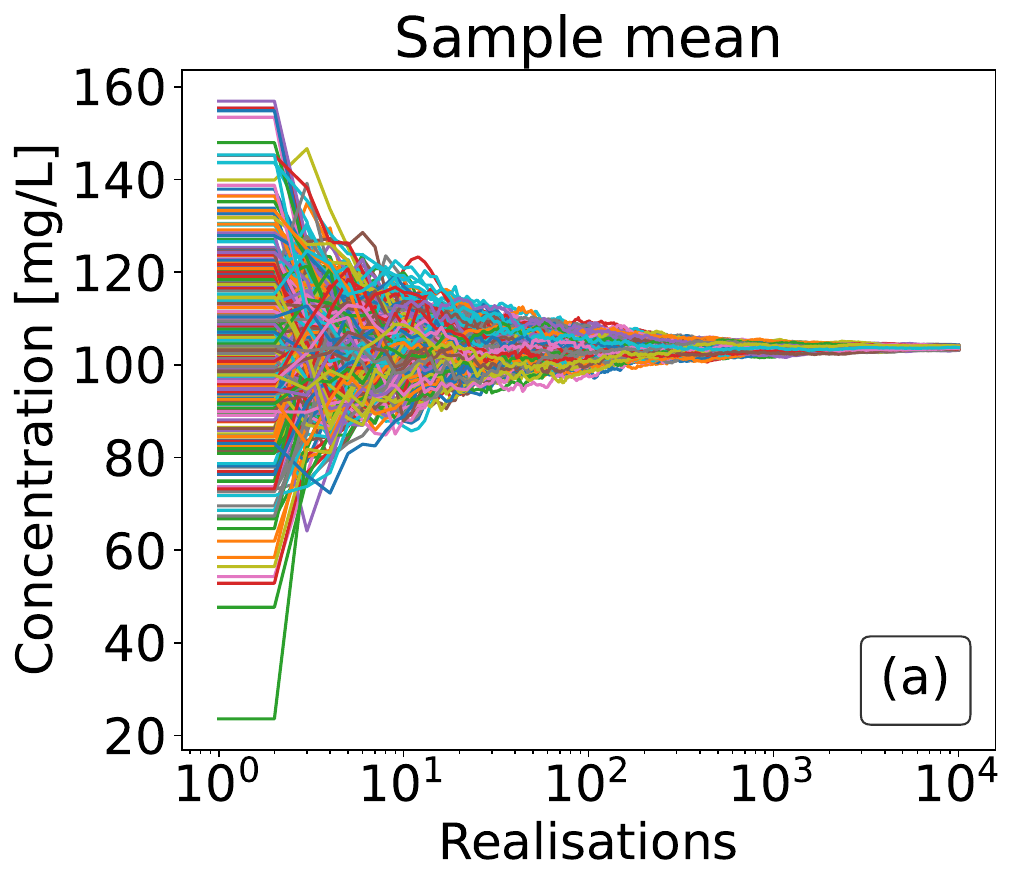}   \hspace{0.5cm}
    \includegraphics[scale=0.26]{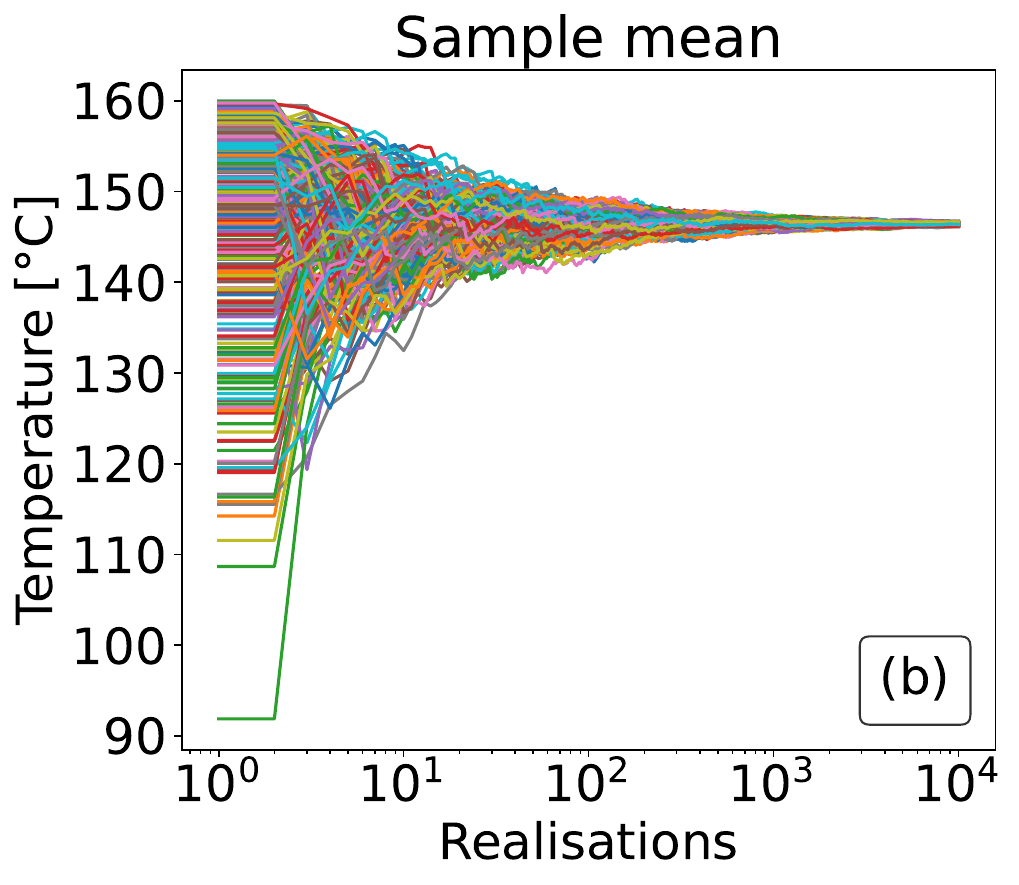} \\
    
    \includegraphics[scale=0.26]{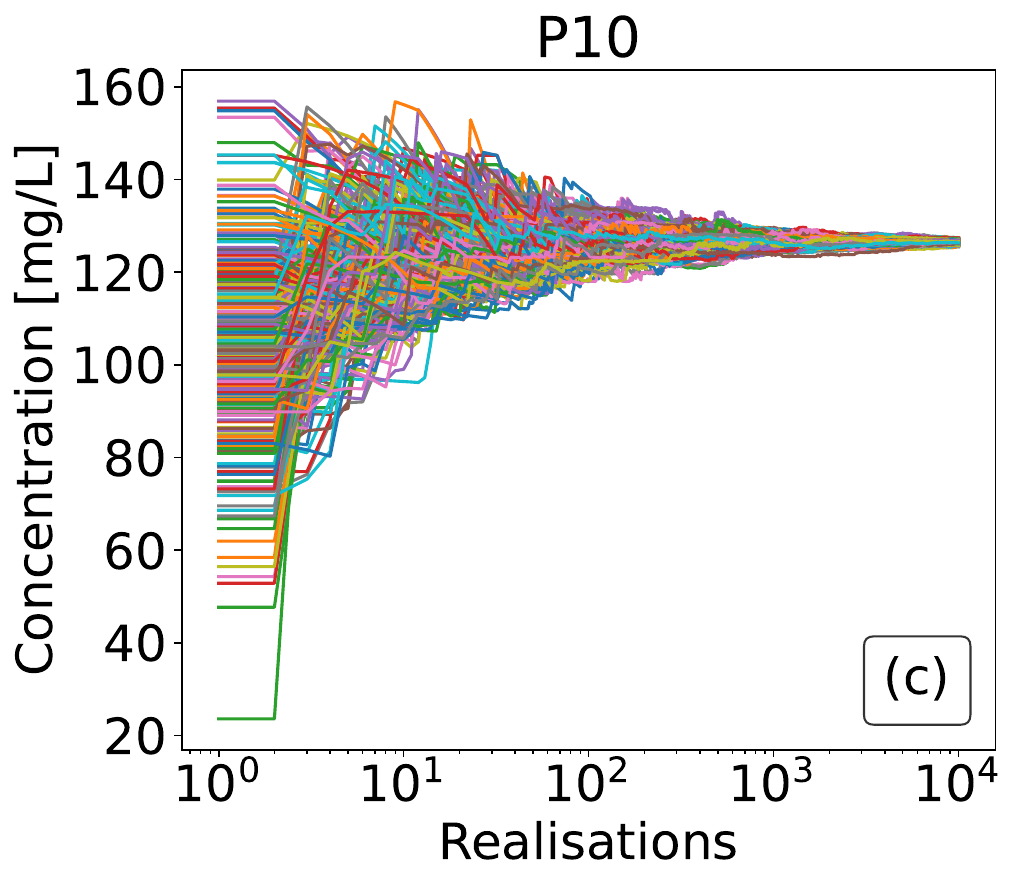} \hspace{0.5cm}
    \includegraphics[scale=0.26]{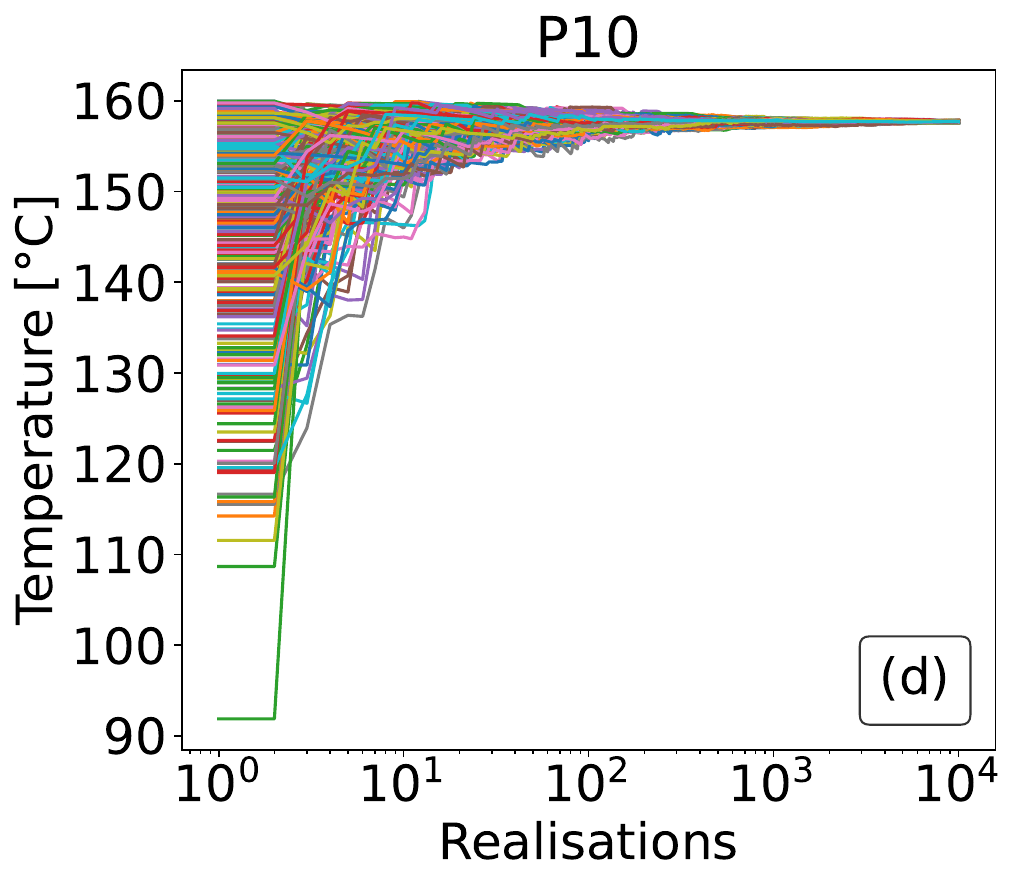} \\
    
    \includegraphics[scale=0.26]{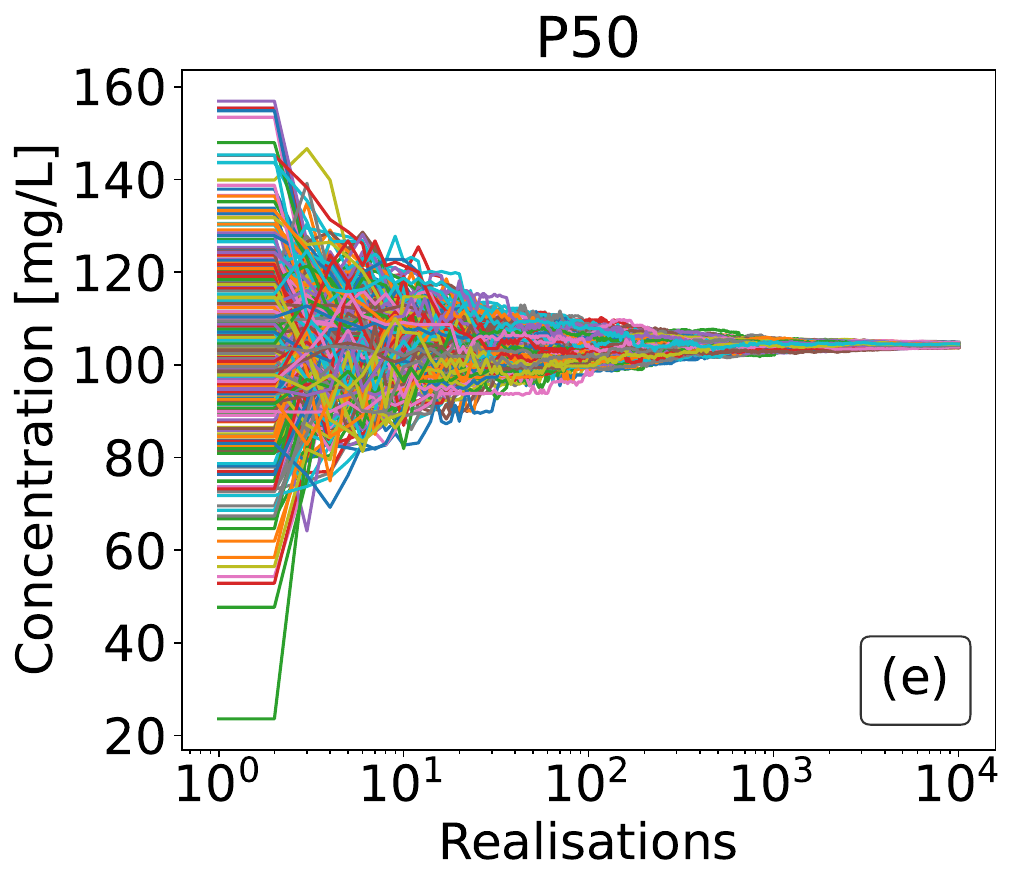} \hspace{0.5cm}
    \includegraphics[scale=0.26]{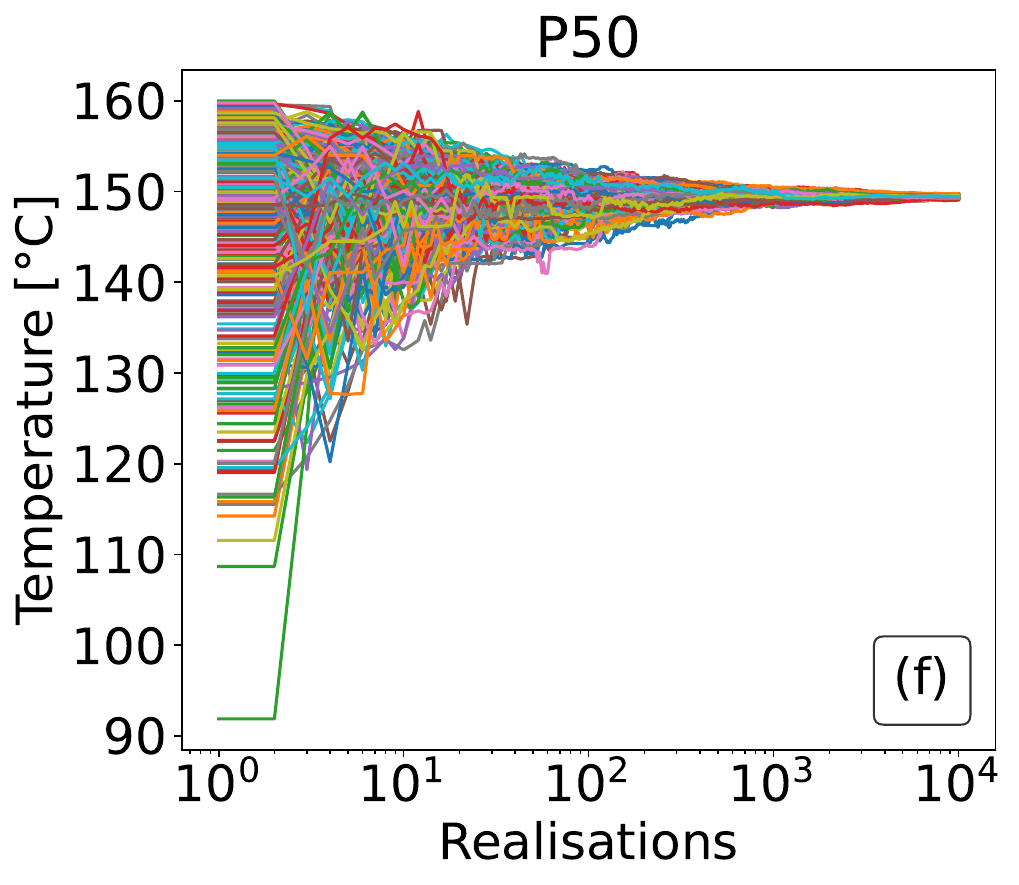} \\
    
    \includegraphics[scale=0.26]{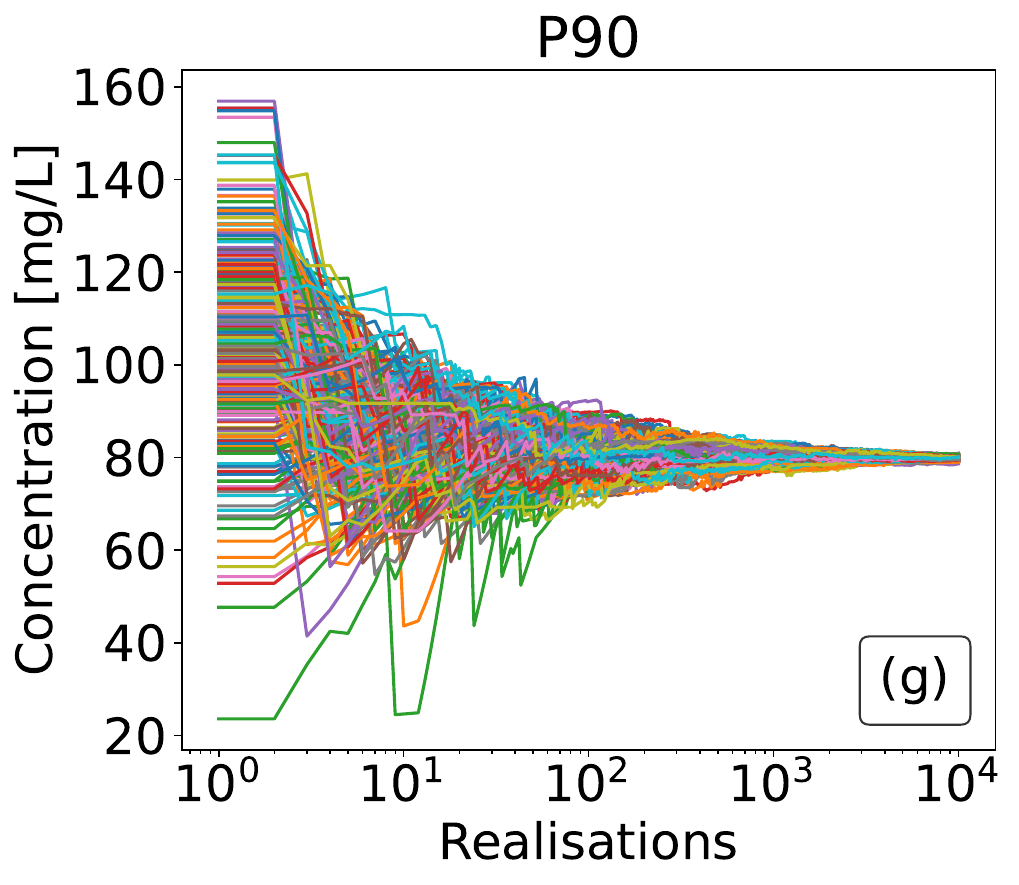} \hspace{0.5cm}
    \includegraphics[scale=0.26]{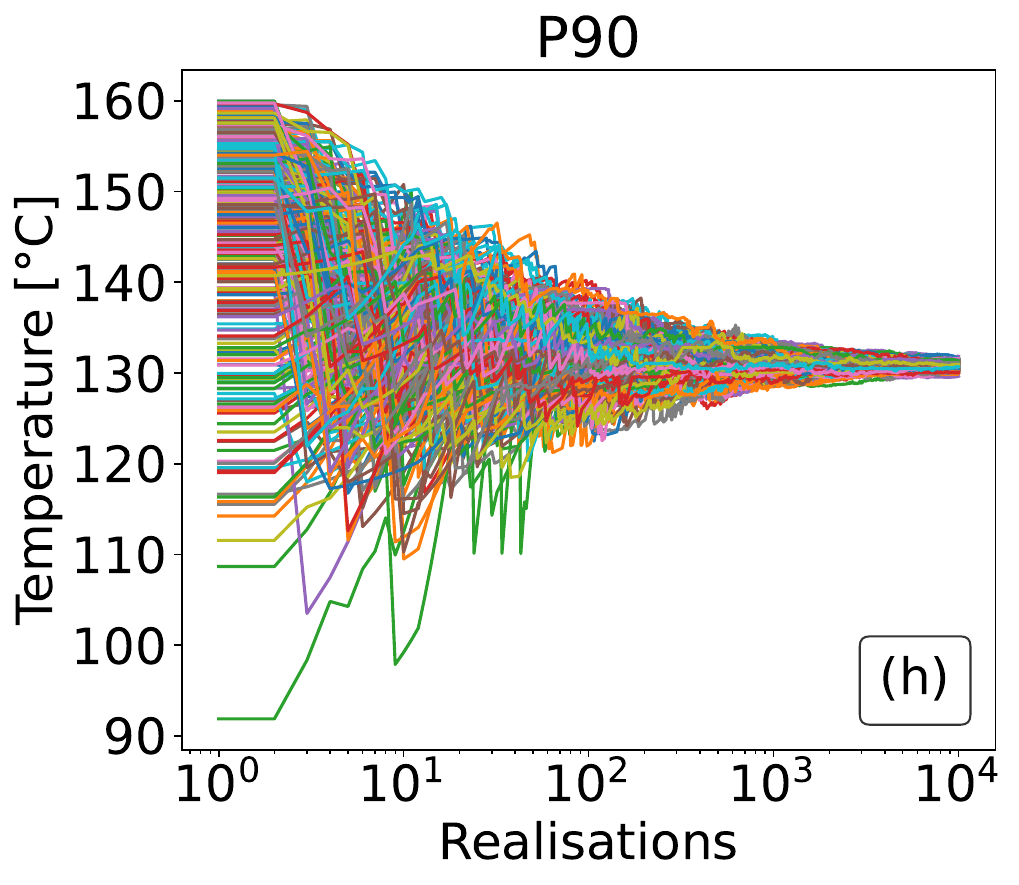} \\   
    
    \caption{The convergence of the Monte Carlo simulations for the network with $30$ fractures. The left column is the concentration, and the right column is the temperature. The rows are the sample mean ((a), (b)), P$10$ ((c), (d)), P$50$ ((e), (f)), and P$90$ ((g), (h)).  
    }
    \label{fig:mc_conv_density_30}
\end{figure}

Fig. \ref{fig:mc_conv_density_30} shows the sample mean, P$10$, P$50$ and P$90$ of the extracted concentration and temperature after $60$ years as a function of the realisations for the density with $30$ fractures. 
The curves in Fig. \ref{fig:mc_conv_density_30}a are constructed as follows. 
We calculate the sample mean of the dataset that contains the $n$ first realised concentrations after $60$ years, $n=1,\, 2,\, 3, \,\hdots, 10,000$. The sample mean of the concentration is recalculated with the increasing number of realised concentrations. When this step is done, we resample the realised concentrations and calculate the sample mean of the resampled concentrations in the same fashion. The reshuffling procedure is repeated $300$ times, and Fig. \ref{fig:mc_conv_density_30}a shows everything put together. 
Fig. \ref{fig:mc_conv_density_30}b demonstrates the outcome of this procedure applied to the temperature. Figs \ref{fig:mc_conv_density_30}c-h show the procedure applied on P$10$, P$50$ and P$90$. A clear convergence is observed for all the plotted quantities. We observed a similar convergence for the networks with $8$ and $82$ fractures.

\end{appendices}

\end{document}